\documentclass[fleqn,usenatbib]{mnras}

\usepackage{bm}
\usepackage[T1]{fontenc}

\DeclareRobustCommand{\VAN}[3]{#2}
\let\VANthebibliography\thebibliography
\def\thebibliography{\DeclareRobustCommand{\VAN}[3]{##3}\VANthebibliography}

\usepackage{graphicx}	
\usepackage{amsmath}	

\newcommand\dd{{\mathrm d}}
\newcommand\ii{{\mathrm i}}
\newcommand\vece{{\bmath e}}
\newcommand\vecH{{\bmath H}}
\newcommand\vecl{{\bmath l}}

\newcommand\vecN{{\bmath N}}
\newcommand\vecx{{\bmath x}}
\newcommand\vecX{{\bmath X}}
\newcommand\bcdot{{\bmath\cdot}}
\newcommand\btimes{{\bmath\times}}

\usepackage{newtxtext,newtxmath}

\title[Nonlinear  warped discs]{Nonlinear behaviour of warped discs around a central object with a quadrupole moment}

\author[Deng \& Ogilvie]{
Hongping Deng $^{1}$\thanks{E-mail:hpdeng353@gmail.com}
and Gordon I. Ogilvie $^{1}$\thanks{gio10@cam.ac.uk} \\
$^{1}$ Department of Applied Mathematics and Theoretical Physics, University of Cambridge, \\ Centre for Mathematical Sciences, Wilberforce Road, Cambridge CB3 0WA, UK}

\begin{document}
\label{firstpage}
\pagerange{\pageref{firstpage}--\pageref{lastpage}}
\maketitle

\begin{abstract}
The nonlinear behaviour of low-viscosity warped discs is poorly understood. We verified a nonlinear bending-wave theory, in which fluid columns undergo affine transformations, with direct 3D hydrodynamical simulations. We employed a second-order Godunov-type scheme, Meshless Finite Mass (MFM), and also the Smoothed Particle Hydrodynamics (SPH) method, with up to 128M particles. For moderate nonlinearity, MFM maintains well the steady nonlinear warp predicted by the affine model for a tilted inviscid disc around a central object with a quadrupole moment. However, numerical dissipation in SPH is so severe that even a low-amplitude nonlinear warp degrades at a resolution where MFM performs well. A low-amplitude arbitrary warp tends to evolve towards a nonlinear steady state. However, no such state exists in our thin disc with an angular semi-thickness $H/R=0.02$ when the outer tilt angle is beyond about $14^\circ$. The warp breaks tenuously and reconnects in adiabatic simulations, or breaks into distinct annuli in isothermal simulations. The breaking radius lies close to the location with the most extreme nonlinear deformation. Parametric instability is captured only in our highest-resolution simulation, leading to ring structures that may serve as incubators for planets around binaries.
\end{abstract}




\section{Introduction}
Astrophysical discs around massive bodies are ubiquitous and sometimes non-planar or warped. Warped discs exist on a wide range of scales. They are observed, for example, in maser emission around supermassive black holes in the centres of active galaxies such as NGC 4258~\citep{Miyoshi1995}. Warped circumstellar discs are revealed by dust emission around the protostar IRAS 04368+2557~\citep{Sakai2019} and in the triple star system GW Ori~\citep{Kraus2020, Bi2020, Smallwood2021}. In addition, there are numerous indirect observations of warps such as shadows in outer regions of protoplanetary discs cast by inner warps~\citep[see, e.g.,][]{Marino2015, Stolker2016, Casassus2018, Benisty2018, Muro2020} and super-orbital modulations of many X-ray binaries~\citep{Ogilvie2001, Kotze2012}.

Warps can induce oscillatory horizontal flows whose amplitude is often thought to be limited by the effective viscosity of the disc~\citep{Papaloizou1983}. When the dimensionless viscosity parameter $\alpha$~\citep{Shakura1973} is large compared to the disc's angular semi-thickness $H/R$, where $H$ is the scale-height at radius $R$, the warp evolves diffusively, as first shown in a linear analysis by \cite{Papaloizou1983} and then in a fully nonlinear derivation by~\cite{Ogilvie1999}. \cite{Ogilvie1999} presented general formulae for the torques associated with the internal flow and also identified a  precessional torque that must be added to the heuristic diffusion equations of \cite{Pringle1992}. Three-dimensional hydrodynamic simulations utilizing the Smoothed Particle Hydrodynamics (SPH) scheme~\citep{Lodato2010} found good agreement
with the nonlinear diffusive theory of~\cite{Ogilvie1999}.

When the disc is nearly inviscid ($\alpha \ll H/R$), however, a low-amplitude warp propagates as a non-dispersive bending wave in a Keplerian disc~\citep{Papaloizou1995, Lubow2000} and as a dispersive wave in a non-Keplerian disc~\citep{Ogilvie1999}. In addition, the oscillatory horizontal flow can be unstable owing to a parametric resonance of inertial waves, as first shown for Keplerian discs by~\cite{Gammie2000}. \cite{Ogilvie2013a,Ogilvie2013b} developed a local warped shearing-box model for slowly evolving warps and showed generic parametric instability for non-Keplerian warped discs. Dedicated simulations verified the widespread parametric instability in the local approximation~\citep{Paardekooper2019} and in a global setup allowing warp propagation~\citep{Deng2021}.

Although some attempts were made to bridge the nonlinear diffusive theory and the linear bending-wave theory \citep{Martin2019, Dullemond2021}, a nonlinear theory for inviscid warped discs is not well established yet. \cite{Ogilvie2006} derived a dispersive wave equation for weakly nonlinear bending waves in Keplerian discs by asymptotic analysis and found that nonlinearity tended to broaden the waves in physically realistic cases. More recently, \cite{Ogilvie2018} developed an affine model for thin astrophysical discs beyond the equations of 2D hydrodynamics. The disc is thought of as a set of fluid columns each undergoing an affine transformation in three dimensions, consisting of a translation plus a linear transformation. The time-evolution of the two deformations is derived using Hamilton's Principle for a 3D fluid. The affine model provides a good description of long-wavelength modes in a 3D polytropic disc and reproduces precisely the linear theory of 3D warped discs. Being free from any small-amplitude assumptions, the affine model can be a good extension for inviscid warp theories into the nonlinear regime. However, it is unclear how accurately fluid columns undergo affine transformations in nonlinear warps. \cite{Fairbairn2021a,Fairbairn2021b} developed a local nonlinear ring model to study the resonant behaviour of fluids in warped Keplerian discs; this is equivalent to the affine model in the context of a ring, and similarly ignores complex internal flows such as those resulting from parametric instability. Notably, \cite{Fairbairn2021b} found strong vertical expansion and compression twice per orbit in warped Keplerian discs.

Here we examine the applicability of the affine model as a nonlinear warp theory for inviscid discs. To this end, we focus on simple steady nonlinear warp solutions around a central object with a quadrupolar potential component, and we run 3D hydrodynamical simulations to verify the solutions. Owing to the Lagrangian nature of the affine model, we chose a second-order Godunov-type Lagrangian scheme, the Meshless Finite Mass method~\citep[MFM; ][]{Hopkins2015}, and also the Smoothed Particle Hydrodynamics method~\citep[SPH; for recent reviews see][]{Price2012,Rosswog2015} for comparison. 

We will see below that 3D simulations adequately verify the affine model, although numerical dissipation can degrade the assumed steady state, especially in SPH simulations. We report warp breaking once the outer tilt angle goes beyond a critical maximum tilt predicted by the nonlinear theory. In adiabatic simulations, the broken disc reconnects instead of evolving into distinct annuli in isothermal simulations \citep[see, e.g.,][]{Drewes2021}. Parametric instability develops only in the highest-resolution simulation so far, employing 128M particles, resulting in ring structures in the disc.

The remainder of this paper is structured as follows. We first review both linear and nonlinear theories of warped inviscid discs in Section \ref{sec:theory} and then apply the theories to find steady warp solutions around a quadrupole in Section \ref{sec:NLS}. In Sections \ref{sec:IC} and \ref{sec:results}, we construct numerical representations of the steady warp solutions and evolve these models to test the warped disc theories. We draw conclusions in Section \ref{sec:con}.

\section{Theories for inviscid warped discs}
\label{sec:theory}

The evolution of small-amplitude warps has been treated by \cite{Papaloizou1995}, and a dispersive wave equation for bending waves in inviscid non-Keplerian discs can be found in \cite{Ogilvie1999}. The warped disc can be thought of as a continuous set of circular rings whose normal vectors, $\bm{l}(R,t)$, vary with the radius $R$ of the ring and time $t$. The governing equations \citep{Lubow2000} for small-amplitude warps in a nearly Keplerian, non-self-gravitating disc of small viscosity ($\alpha\ll1$) read
\begin{align}
  &\Sigma R^2 \Omega \left( \frac{\partial W}{\partial t} + \frac{\eta^2-\Omega^2}{\Omega^2} \frac{i\Omega}{2}W\right)=\frac{1}{R}\frac{\partial \mathcal{G}}{\partial R}, \label{eq:leq1}\\                &\frac{\partial\mathcal{G}}{\partial t}+\frac{\kappa^2-\Omega^2}{\Omega^2}\frac{i\Omega}{2}{\mathcal{G}}+\alpha \Omega \mathcal{G}=\frac{1}{4}\Sigma H^2R^3\Omega^3\frac{\partial W}{\partial R}, \label{eq:leq2}
\end{align}
where $\Sigma(R)$ is the surface density and 
$H(R)$ is the scale-height.
Here $W(R,t)=l_x + il_y$ is the complex tilt variable and $\mathcal{G}(R,t)=G_x+iG_y$ is the complex internal torque variable, while $\Omega(R)$, $\kappa(R)$ and $\eta(R)$ are the orbital, epicyclic and vertical oscillation frequencies, respectively.
The torque $\mathcal{G}$ is associated with an internal flow, driven by pressure gradients in the warped disc. In a frame rotating with the fluid, this flow corresponds to an epicyclic oscillation that is forced at its natural frequency in a Keplerian disc with a stationary warp; equation~(\ref{eq:leq2}) allows the response to this resonant forcing to be moderated either by viscous damping or by detuning of the resonance resulting from non-Keplerian rotation or time-dependence of the warp. In a non-rotating frame, the streamlines of the disc acquire an elliptical form, with eccentricity proportional to the height above the midplane \citep[see also][]{Deng2021}.

The above equations derived from asymptotic analysis are not easily generalized to nonlinear warps. However, the affine model proposed recently by \cite{Ogilvie2018} reduces to the above equations in the limiting case of small-amplitude warps and could be useful for general warped discs. The disc is viewed as a set of fluid columns that are vertical in the reference state (an axisymmetric, unwarped disc). The centre of mass of each column, located at $\bm{x}_0$ in the reference state, moves to $\bm{X}(\bm{x}_0,t)$ in the dynamical (warped) state. The 3D fluid elements within each column are further identified by the dimensionless label
\begin{equation}
\zeta=\frac{z_0}{H_0},
\end{equation}
where $z_0$ and $H_0$ are the vertical coordinate and scale-height in the reference state.
In the dynamical state, the position vector of a fluid element is $\bm{x}=\bm{X}+\bm{H}\zeta$.
\cite{Ogilvie2018} derived the governing equations for the variables $\bm{X}$ and $\bm{H}$ in this affine transformation using Hamilton's Principle for a 3D fluid and made a correspondence with existing theories of warped discs. The affine model is capable of describing both the warping of the midplane, through the motion of the column centres described by $\bm{X}$, and also the internal flows leading to the torque that transmits the warp, through a tilting of the columns described by $\bm{H}$.

\section{Steady warps around a quadrupole}
\label{sec:NLS}

\begin{figure*}
  \includegraphics[width=\linewidth]{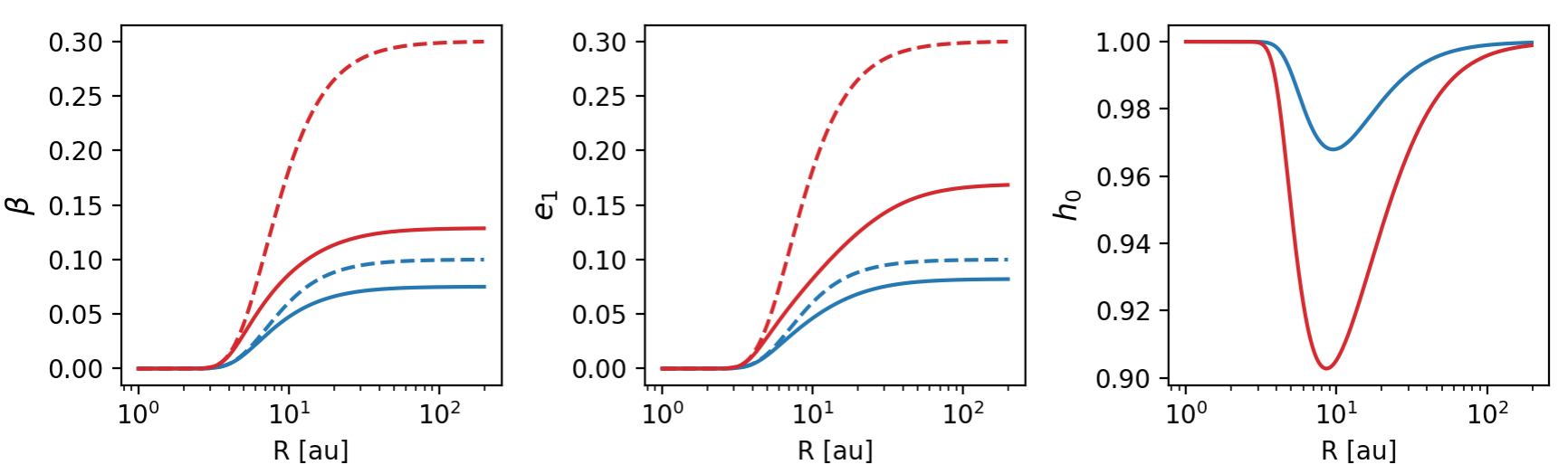}
  \caption{Steady warp solutions to the linear inviscid warp equations (dashed lines) and the nonlinear affine model (solid lines). The tilt angle $\beta$ is measured in radians. For the affine model, $e_1$ is the eccentricity at one scale-height above the mid-plane and $h_0$ is the vertical rescaling factor at $\phi=0$ (see text).  The nonlinear solutions connect to the linear ones at small radii where $\beta\ll1$. We henceforth refer to the low- and high-amplitude nonlinear solutions as NS1 and NS2, respectively.}
  \label{fig:NLS}
\end{figure*}

The two sets of equations described above are difficult to solve
in general. Therefore we consider a \emph{steady} warp in the gravitational potential
\begin{equation}
\Phi=-\frac{GM}{\sqrt{R^2+z^2}}\left [1+\frac{Q^{2}(R^2-2z^2)}{3(R^2+z^2)^2}\right ],\label{eq:phi}
\end{equation}
which is generated by a monopole of mass $M$ plus an axisymmetric quadrupole aligned with the $z$-axis. If the central object is a binary of total mass $M=M_1+M_2$ and semimajor axis $a$ in the plane $z=0$, then the strength of the quadrupole is given by $Q^2=3M_1M_2a^2/4M^2$. 
Circular particle orbits in the mid-plane have 
\begin{align}
&  \Omega^2=\frac{GM}{R^3}\left ( 1+ \frac{Q^2}{R^2}\right ), \\
&  \kappa^2=\frac{GM}{R^3}\left ( 1- \frac{Q^2}{R^2}\right ), \\
&  \eta^2 =\frac{GM}{R^3}\left ( 1+ \frac{3Q^2}{R^2}\right ).
\end{align}
The orbits for $R<Q$ are unstable ($\kappa^2<0$). For $R\gg Q$ the apsidal precession frequency is given by
\begin{equation}
\Omega - \kappa \approx \omega_\text{p}, \quad \omega_\text{p}=Q^2\left ( \frac{GM}{R^7}\right )^{1/2},
\end{equation} and the nodal precession frequency is $-\omega_\text{p}$.

Steady warp solutions around such a quadrupole exist in an infinite disc when the internal torque is balanced by the gravitational torque everywhere~\citep[see, e.g.,][]{Facchini2013, Foucart2013}. For ease of numerical simulations, we consider a vertically isothermal disc of constant angular semi-thickness $\epsilon=H_0/R=0.02$ with $\Sigma_0\propto R^{-1}$. We work in the context of circumstellar discs, taking $M=M_\odot$ and $Q=1$ au, but we note that the problem we study is scale-free. We used a unit system of 1 au, 1 solar mass and $G=1$ in the calculations and simulations reported below. The gas evolves either adiabatically, with $\gamma=5/3$, or nearly isothermally, with $\gamma=1.001$. 

\subsection{Solutions to the linear equations}
For our specific disc model, equations (\ref{eq:leq1}) and (\ref{eq:leq2}) become
\begin{align}
  &\frac{\partial W}{\partial t} = -i\omega_\text{p}W + \frac{1}{\Sigma R^3\Omega}\frac{\partial \mathcal{G}}{\partial R},\label{dwdt}\\
  & \frac{\partial\mathcal{G}}{\partial t}=i\omega_\text{p} \mathcal{G}+\frac{1}{4}\Sigma H^2R^3\Omega^3\frac{\partial W}{\partial R}.\label{dgdt}
\end{align}
If we regard $\Omega \propto R^{-3/2}$ as Keplerian in these equations, which is a good approximation for $R\gg Q$, then a solution for a steady-state warp is
\begin{equation}
W=W_\infty \text{exp} \left(-\frac{Q^2}{\epsilon R^2}\right),
\label{eq:LS}
\end{equation}
where $W_\infty$ is the tilt at large distance. Two solutions of the linear bending-wave equations are plotted in the left panel of Fig.~\ref{fig:NLS}. Note the tilt angle $\beta$ is related to the tilt vector by $\bm{l}=(\sin \beta, 0,\cos \beta)$ and for the small $\beta$ considered here $\beta \approx l_x$. In our thin disc model, $\beta$ is negligible for $R<3$ au in Fig.~\ref{fig:NLS}, which in turn validates our approximation of $\Omega$ as Keplerian in obtaining the solutions.

\subsection{Nonlinear warps in the affine model}
\label{sec:nls}
We here outline an application of the affine model \citep{Ogilvie2018} to nonlinear warps around a quadrupole. (Further details of the affine theory of nonlinear warps will be presented elsewhere.) Let us regard the reference state as an axisymmetric, unwarped disc, and use polar coordinates $(R,\phi)$ to label its upright fluid columns. The vertical coordinate of a fluid element is $z_0=H_0\zeta$, where $H_0(R)$ is the scale-height of the reference disc and $\zeta$ is a dimensionless label.

To reach the (untwisted) warped state, each ring of the disc is rotated through an angle $\beta(R)$ about the $y$-axis, so that its unit tilt vector is $\vecl=(\sin\beta,0,\cos\beta)$. This rotation describes the mapping of the column centres, $\vecX(R,\phi)$, in the affine model. In general, fluid elements have position vectors $\vecx=\vecX+\vecH\zeta$, where $\vecH$ is the scale vector of each column and the Lagrangian coordinate $\zeta$ is preserved from the reference state. The Jacobian of the transformation from the reference state to the warped disc is the triple scalar product
\begin{equation}
  J=\frac{1}{R}\left(\frac{\partial\vecx}{\partial R}\btimes\frac{\partial\vecx}{\partial\phi}\right)\bcdot\frac{\partial\vecx}{\partial z_0}.
\end{equation}
The essential approximation in the affine model, which is accurate for large-scale deformations of thin discs, is to regard $J$ as independent of $\zeta$, using its value at $\zeta=0$. This gives $J=\vecH\bcdot\vecN/H_0$, where
\begin{equation}
  \vecN=\frac{1}{R}\left(\frac{\partial\vecX}{\partial R}\btimes\frac{\partial\vecX}{\partial\phi}\right)
\end{equation}
is a dimensionless vector normal to the warped midplane of the disc. It can be evaluated for an untwisted warp as $\vecN=\vecl+\psi\cos\phi\,\vece_R$ (in agreement with Appendix A of \citealt{Ogilvie2001}), where $\psi=R\,\dd\beta/\dd R$ is the warp amplitude and $\vece_R=(\cos \beta \cos\phi, \sin \phi, -\sin\beta \cos \phi)$ is the radial unit vector.

The scale vector can be decomposed into `vertical' and `horizontal' components, parallel and perpendicular to $\vecl$. The vertical component corresponds to a variation of the thickness of the disc, which is a nonlinear effect and is described below. The horizontal component captures the tilting of the fluid columns caused by the pressure gradients in the warped disc. As described in Section~\ref{sec:theory}, this aspect of the internal flow in a warped disc, which is present in linear theory, corresponds to an elliptical deformation of the streamlines of the disc, proportional to $\zeta$. If this shearing motion corresponds to a Keplerian eccentricity $e=e_1\zeta$ in phase with the warp (as expected for a stationary warp in an inviscid disc), where $e_1(R)$ is the eccentricity at one scale-height above the midplane, then $\vecH\bcdot\vece_R=-e_1R\cos\phi$. We then have
\begin{equation}
  J=h-f,\qquad
  f=\chi\cos^2\phi,
\label{jhf}
\end{equation}
where $h=\vecH\bcdot\vecl/H_0$ is the factor by which the thickness is changed, while
\begin{equation}
  \chi=\frac{\psi e_1}{H_0/R}
\label{chi}
\end{equation}
involves the product of the warp and the shear, and corresponds to the variable $Z_1$ in \cite{Fairbairn2021b} in the case when this is real.

\begin{figure}
  \centering
  \includegraphics[width=\linewidth]{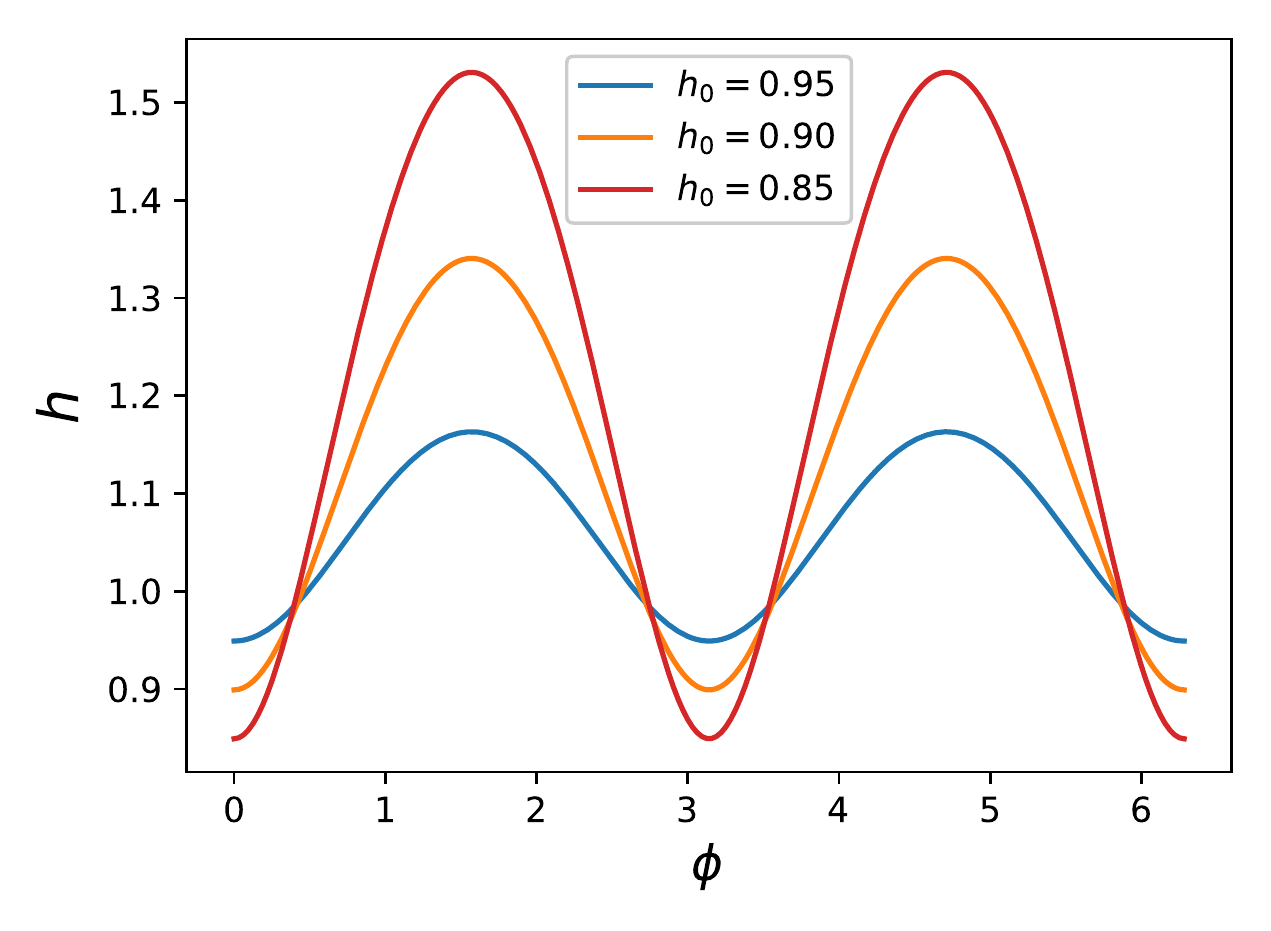}
    \caption{Exemplar vertical rescaling factors for an adiabatic disc ($\gamma=5/3$) as a function of the azimuthal angle, labelled by their minimum value at $\phi=0$. The graphs show the non-hydrostatic oscillation of the scale-height induced by a warp.}
  \label{fig:Hz}
\end{figure}
The Lagrangian of the disc, excluding the integrable part associated with Keplerian orbital motion around a point mass, can be well approximated as
\begin{equation}
  L=\int\left[\frac{1}{2}\Omega^2H_0^2\left(h_\phi^2-h^2\right)-\Phi_Q-\frac{J^{-(\gamma-1)}}{\gamma-1}\frac{P_0}{\Sigma_0}\right]\dd m,
\end{equation}
with contributions from the vertical kinetic energy associated with a varying thickness ($h_\phi$ meaning $\partial h/\partial\phi$), the vertical gravitational  energy due to the central monopole, the gravitational  energy due to the quadrupole (neglecting the thickness of the disc) and the internal energy. Here $\Sigma_0(R)$ and $P_0(R)$ are the vertically integrated density and pressure of the reference disc, respectively, and $\dd m=\Sigma_0\,2\pi R\,\dd R$. Hydrostatic equilibrium of the reference disc implies $P_0=\Sigma_0\Omega^2H_0^2$. The main approximations involved here are that (i) the disc is thin, (ii) the quadrupolar component of the potential is small compared to the monopolar component, (iii) the motion of the disc is well described by affine transformations of fluid columns, and (iv) the warp is of large scale compared to the thickness of the disc. These reasonable assumptions are similar to the approximations made in other analytical treatments of warped discs.

Variation of $L$ with respect to $h$ yields a dynamical equation for the azimuthal variation of the scale-height,
\begin{equation}
  h_{\phi\phi}+h=J^{-\gamma},
\label{hpp}
\end{equation}
which, together with equation~(\ref{jhf}), implies a non-hydrostatic, nonlinear oscillation that is driven at twice the orbital frequency by the geometrical forcing ($f$) due to the combination of warp and shear (see equations \ref{jhf}--\ref{chi}). This is a nonlinear aspect of the dynamics of warped discs that has featured in previous work \citep{Ogilvie1999,Ogilvie2006,Ogilvie2013a,Fairbairn2021b}. We obtain $2\pi$-periodic solutions to these equations by the shooting method, starting with small values of $\chi$ and following the family of solutions quasi-continuously as this parameter is varied. In place of $\chi$, we find it convenient to label the solutions by $h_0$, the minimum value of $h$, which occurs at $\phi=0$. Exemplar solutions are plotted in Fig.~\ref{fig:Hz}, labelled by $h_0$. Solutions with $h_0$ ranging from 0.999 to 0.7 every 0.001 on a uniform $\phi$ grid of interval $\pi/100$ are stored for interpolation in Section~\ref{sec:trans}.

The secular effects of the vertical oscillation and the quadrupolar potential on the warp and tilt can be deduced by orbit-averaging the Lagrangian, which produces
\begin{equation}
  \langle L\rangle=\int\frac{GM}{R^3}\left[H_0^2\bar L(\chi)+\frac{Q^2\left(3\cos^2\beta-1\right)}{6(1-e_1^2)^{3/2}}\right]\dd m,
\end{equation}
where
\begin{equation}
  \bar L(\chi)=\frac{1}{2\pi}\int_0^{2\pi}\left[\frac{1}{2}\left(h_\phi^2-h^2\right)-\frac{J^{-(\gamma-1)}}{\gamma-1}\right]\dd\phi
\end{equation}
is a dimensionless function (depending also on $\gamma$) studied by \citet{Fairbairn2021b}, who called it $\langle L_{10}\rangle(X)$.

Variation of $\langle L\rangle$ with respect to $\beta$ gives
\begin{equation}
  \frac{\dd}{\dd R}\left[\Sigma_0H_0\left(-\frac{\dd\bar L}{\dd \chi}\right)e_1\right]=\frac{\Sigma_0}{R^2}\frac{Q^2\cos\beta\sin\beta}{(1-e_1^2)^{3/2}}
\label{ode1}
\end{equation}
and variation of $\langle L\rangle$ with respect to $e_1$ gives
\begin{equation}
  R^2H_0\left(-\frac{\dd\bar L}{\dd \chi}\right)\frac{\dd\beta}{\dd R}=\frac{Q^2\left(3\cos^2\beta-1\right)e_1}{2(1-e_1^2)^{5/2}}.
\label{ode2}
\end{equation}
We note that $-\dd\bar L/\dd \chi$ is a positive, increasing function of $\chi$ that is defined for all negative $\chi$ but ceases to exist when $\chi$ exceeds a positive critical value, depending on $\gamma$. (In fact, the branch of solutions of the forced nonlinear vertical oscillator folds over at this point and becomes unstable.) We have a nonlinear system of ordinary differential equations for $\beta(R)$ and $e_1(R)$. In the limit $\beta,e_1,\chi\ll1$, in which $-\dd\bar L/\dd \chi$ can be approximated as $1/2$, it reduces to the linear system
\begin{align}
  &\frac{\dd}{\dd R}\left(\frac{1}{2}\Sigma_0H_0e_1\right)=\frac{\Sigma_0Q^2\beta}{R^2},\\
  &\frac{1}{2}R^2H_0\frac{\dd\beta}{\dd R}=Q^2e_1,
\end{align}
which is equivalent to equations (\ref{dwdt})--(\ref{dgdt}) in a steady state, if we identify $W$ with $\beta$ and $\mathcal{G}$ with $\frac{1}{2}\ii GM\Sigma_0H_0e_1$.

Solutions of the coupled nonlinear ordinary differential equations (\ref{ode1}) and (\ref{ode2}) were obtained numerically, starting at small radii where the desired solution is of the same form as in the linear problem. Indeed, the family of solutions can be parametrized by $W_\infty$, the value of $\beta$ that would be obtained at large radii if the warp remained in the linear regime (see equation \ref{eq:LS}). It is found that the outer tilt angle is smaller than $W_\infty$ because of nonlinear effects, principally the nonlinearity of the warp at intermediate radii. Two examples of nonlinear solutions are plotted in Fig.~\ref{fig:NLS}. We found that, for $\epsilon=0.02$ and $\gamma=5/3$, it is not possible to find solutions with outer tilt angles exceeding about $0.2$ radians.

\section{Numerical setup}
\label{sec:IC}

We use primarily the Meshless Finite Mass (MFM) scheme of the multi-method hydrodynamic code GIZMO \citep{Hopkins2015} to simulate the steady warps around a central object with a quadrupole moment, defined by equation~(\ref{eq:phi}). In addition, we provide several SPH simulations with the GIZMO code for comparison.  The MFM and SPH methods share the same algorithm for gravity and computational domain decomposition and thus any discrepancy between the results merely reflects the difference in the hydrodynamic methods \citep [see, e.g.,][]{Deng2017}.

Simulations of \emph{inviscid} warped discs with SPH are complicated by the artificial viscosity, necessary for shock capturing, which may lead to unwanted numerical dissipation, especially in strongly warped regions \citep{Drewes2021}. The numerical dissipation is sensitive to the implementation of the artificial viscosity \citep[see, e.g.,][]{Cullen2010, Hopkins2015, Wadsley2017, Price2018}. However, the effects of various artificial viscosity implementations on warped disc simulations have not been explored systematically. Here the artificial viscosity in GIZMO SPH~\citep{Hopkins2015} stems from a variant proposed by~\cite{Cullen2010}. 

The MFM method involving no artificial viscosity is believed to be less numerically viscous than SPH. \cite{Deng2021} achieved a numerical $\alpha$ less than 0.001 in warped disc simulations at a mid-plane resolution of $H_0/8$ and captured the parametric instability in a freely evolving warped disc. We used the Wendland C4 kernel for MFM and SPH to maintain well-ordered particles and reduce numerical noise~\citep{Deng2019,Rosswog2015}.

\subsection{The unwarped reference state}
\begin{figure}
  \centering
  \includegraphics[width=1.1\columnwidth]{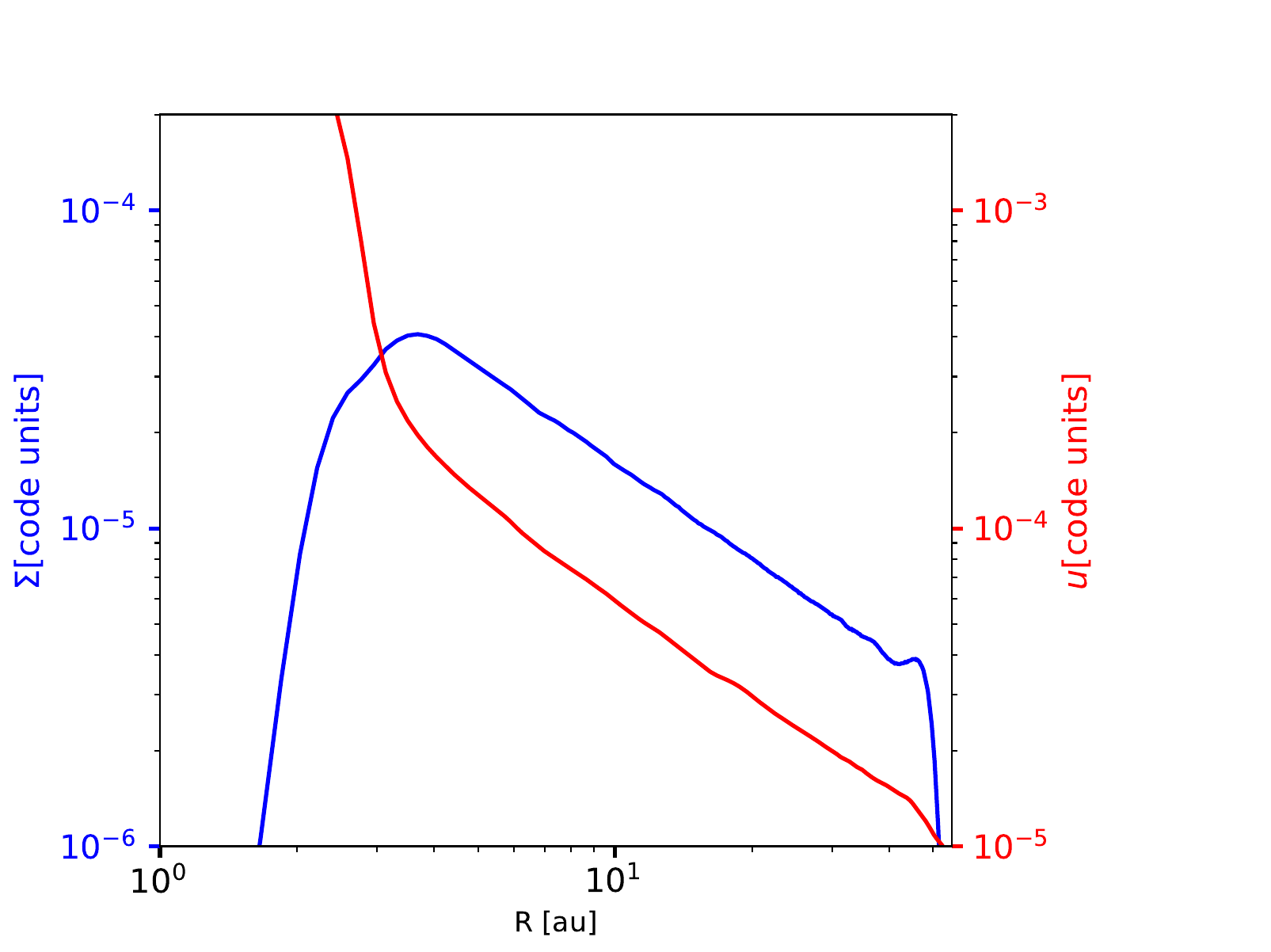}
    \caption{The surface density and internal energy profiles of the unwarped reference state. The corruption of the power law profiles near the inner edge and in the diffusive tail at the outer edge causes little trouble for modelling the steady states in Fig.~\ref{fig:NLS} as these regions are almost flat.}
    \label{fig:IC}
\end{figure}

It is not trivial to generate a particle representation of the $\Sigma_0 \propto R^{-1}$, $H_0/R=\epsilon=0.02$ profile around a central object with a quadrupole moment. In particular, we do not have full control over the innermost unstable region ($R<Q=1$~au). We sample such a density profile in vertical hydrostatic equilibrium spanning 2--50 au with 8M particles via rejection sampling. The total disc mass is 0.05$M_\odot$, although the disc's self-gravity is ignored in this paper. The particles have initial velocities $v_\phi=R\Omega$ and follow a specific internal energy profile $u=\epsilon^2R^2\Omega^2/(\gamma-1)$. We note that the initial velocity is slightly away from radial force balance \citep[see equation 124 in][]{Ogilvie2018}. Nevertheless, the random noise in the particle representation easily overwhelms the velocity errors of order $\epsilon^2 v_\phi$. 

To reduce the particle noise, we relax the initial condition for four outer rotational periods (ORPs) at 49 au by damping the vertical and radial motion every timestep to reach a glass state. We remove particles entering the $R<0.5$ au region. After the relaxation, an equilibrium state is established, as shown in Fig.~\ref{fig:IC}. We further confirm that the disc can maintain this state for $>2$ ORPs, comparable to the warped disc simulation time in Section~\ref{sec:results}. The disc deviates significantly from the desired power-law profile near the inner region of unstable orbits. However, the steady warp solution is flat in this region (see Fig.~\ref{fig:NLS}), so we should still sustain a steady warp in the disc trunk. In principle, we can taper the density profile near the boundaries to avoid violent rearrangements during relaxation \citep[see, e.g.,][]{Lodato2010,Drewes2021}. However, in practice, we find that a wider range $\Sigma_0 \propto R^{-1} $ profile is established in this way compared to relaxation from tapered density profiles.

Finally, for our disc parameters, the mid-plane resolution scale $\delta$ \citep{Deng2019} has the scaling $\delta/H_0 \propto R^{-1/3}$ so that the inner disc is relatively poorly resolved. In the 8M particle reference disc,  $\delta=0.63H_0$ in the mid-plane at 4 au. We split the particles in the quasi-equilibrium state two or four times and relax the resultant discs for 1 ORP to get higher-resolution reference discs. In the 128M particle model, we have $\delta=0.25H_0$ at 4 au; at this resolution, the parametric instability can be marginally captured, as shown by \cite{Deng2021}.  A thicker disc, such as those in \cite{Deng2021}, can be better resolved given the same total number of particles but the steady warp would start at a smaller radius (equation \ref{eq:LS}) where the assumed power-law disc profiles are corrupted (see Fig.~\ref{fig:IC}).

 \subsection{Linear warp models}
We transform the reference disc to the $steady$ warped discs illustrtated in Fig.~\ref{fig:NLS}. The solutions in Section~\ref{sec:NLS} are calculated for an infinite disc, and we shall see later that the outermost part of the disc precesses as a result of the truncation of the disc. The linearised inviscid warp equations and solutions are derived from an Eulerian point of view, so we first apply to the reference disc the Eulerian velocity perturbations due to the inclination of the orbital motion \citep{Lubow2000},
\begin{align}
&  \delta v_\phi=\frac{1}{2}R\Omega \frac{z}{H_0}l_x e^{-i\phi},\\
&  \delta v_r=iR\Omega \frac{z}{H_0}l_x e^{-i\phi}.   \label{eq:vr}
\end{align}
We then apply the rotation matrix
\begin{equation}
\begin{pmatrix}
\cos\beta & 0 & \sin\beta\\
0 & 1 & 0\\
-\sin\beta & 0 & \cos\beta
\end{pmatrix}
\end{equation}to particle positions and velocities. Here $\beta$ for each particle is calculated using equation \ref{eq:LS}.

\begin{figure}
  \centering
  \includegraphics[width=\columnwidth]{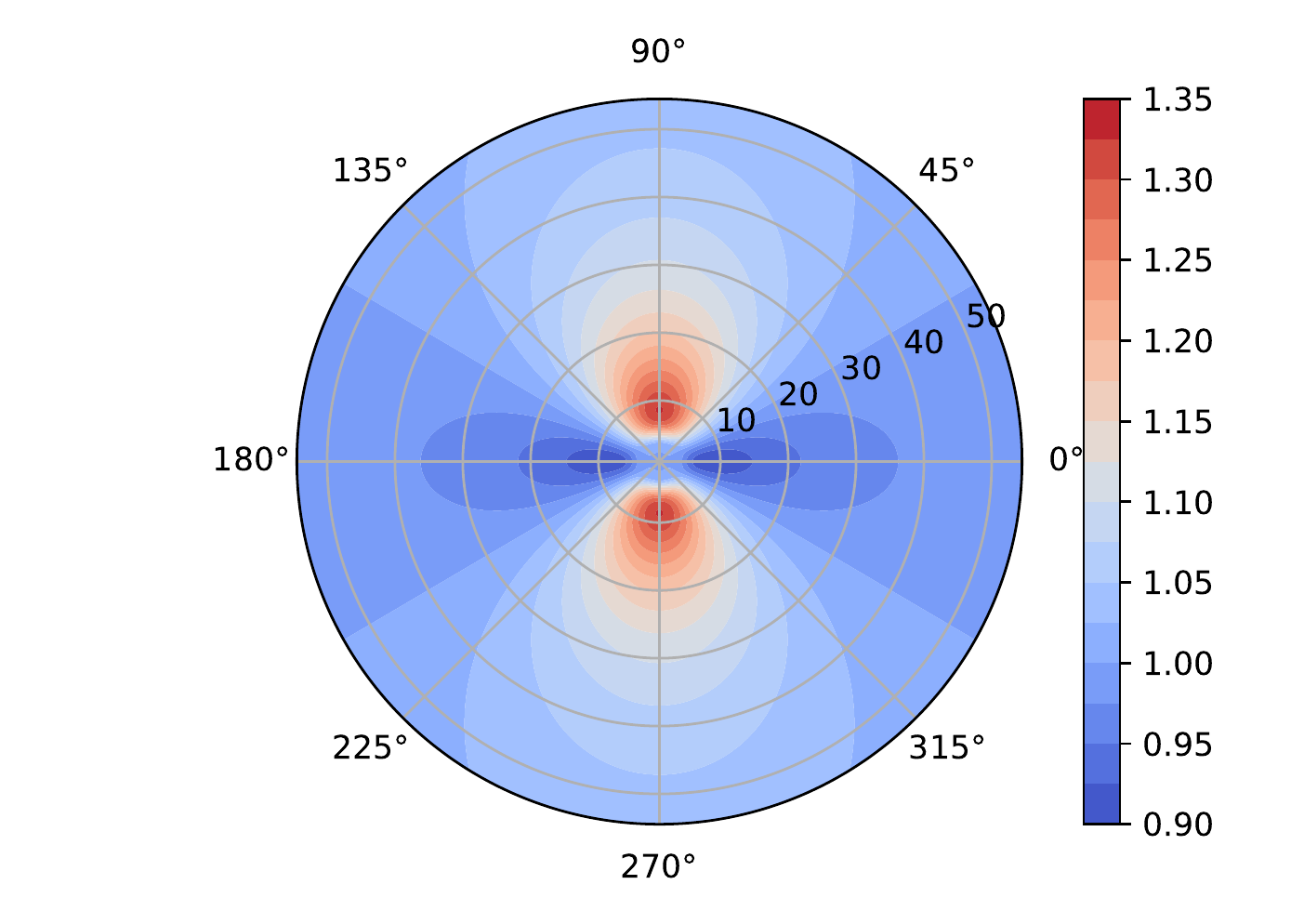}
  \caption{The vertical rescaling factor $h$ for the high-amplitude nonlinear warp (NS2). Compared to the hydrostatic, unwarped reference model, the warped disc is vertically compressed on the $x$-axis and vertically expanded on the $y$-axis.}
  \label{fig:H03}
\end{figure}

\subsection{Nonlinear warp models}
\label{sec:trans}

For the affine model, some care is needed. We first apply the vertical rescaling of the fluid columns, then rotate them. The nonlinear solutions in Fig.~\ref{fig:NLS}, computed by solving equations (\ref{ode1}) and (\ref{ode2}), are quantified by evaluating $\beta$, $e_1$ and $h_0$ at an array of different radii $R_i$ of length $N$; we have about 400 radial grid points in NS1 and NS2. At each $R_i$, we use the known value of $h_0$ to determine the azimuthal variation of the scale-height by interpolating the solutions of equation~(\ref{hpp}) as described in Section~\ref{sec:nls}. We thereby obtain a matrix of $N\times 200$ elements for the vertical scaling factor $h(R,\phi)$. As an example, we plot the rescaling matrix for the high-amplitude nonlinear warp in Fig.~\ref{fig:H03}. We recall that $h$ is the factor by which the dynamical scale-height of the warped disc differs from the hydrostatic scale-height of the unwarped reference model. The disc is vertically compressed by a factor of 0.9 or expanded by a factor of 1.3 around $R=10$ au. We then divide the reference disc into $200N$ columns according to the radial and azimuthal grids. The vertical locations of particles in each column are rescaled by a factor of $h$ calculated by double linear interpolation on the grid.

\begin{figure}
  \centering
  \includegraphics[width=\columnwidth]{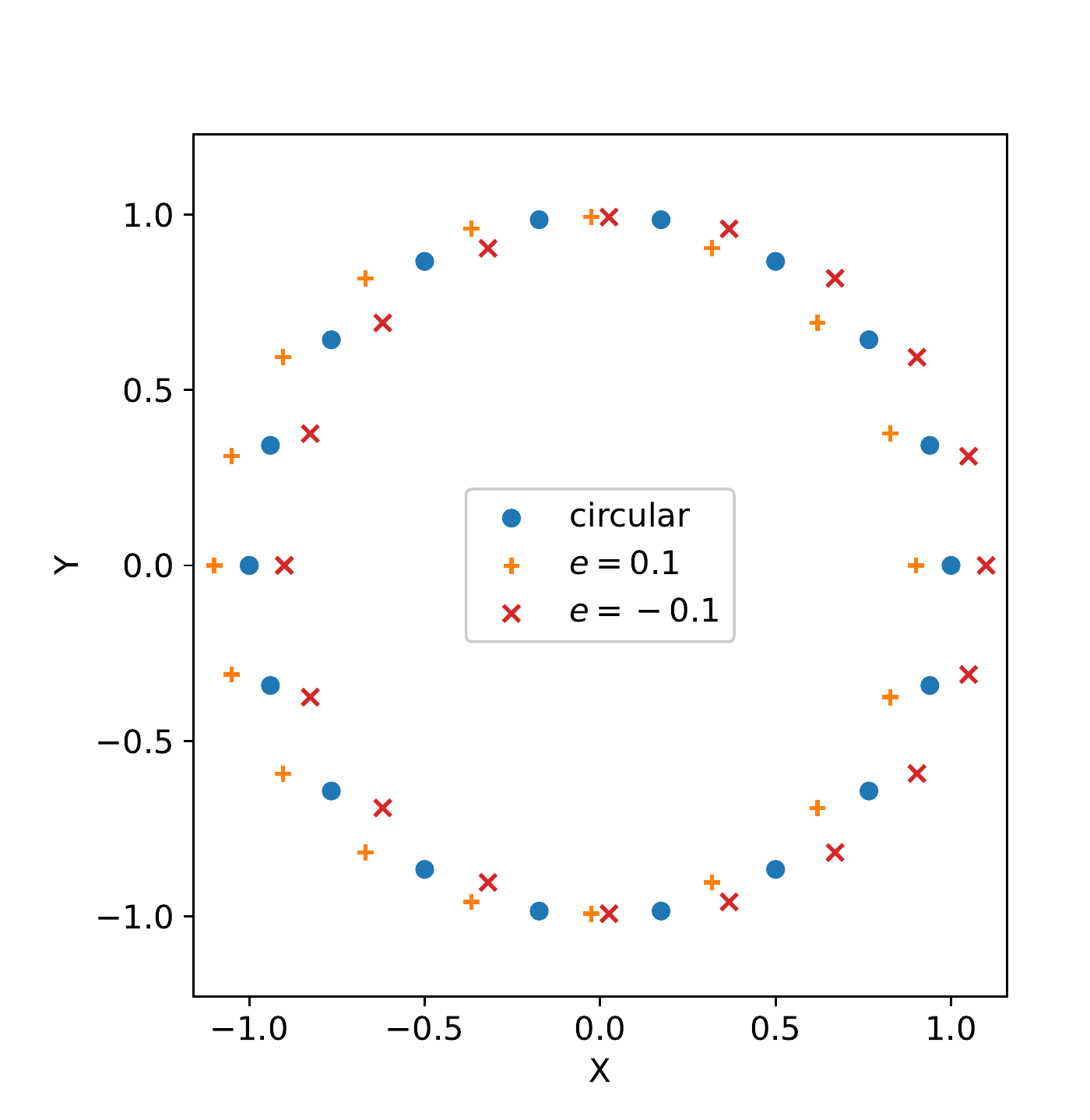}
  \caption{Mapping fluid elements originally on circular orbits to eccentric orbits.}
  \label{fig:orbits}
\end{figure}

In the horizontal plane, particles are remapped onto eccentric orbits directly. The particle semi-major axis is held unchanged, and the eccentricity scales linearly with $\zeta$, i.e.
\begin{equation}
e=e_1\zeta,
\end{equation}where $e_1$ is the characteristic eccentricity at $\zeta=1$ (one scale-height above the midplane).  We note that the $\zeta$ label remains unchanged during the above vertical rescaling. The values of $e_1$ for particles at different radii are calculated via interpolation using the nonlinear bending-wave solutions in Fig.~\ref{fig:NLS}. The true anomaly ($\nu$) is approximated from the mean anomaly ($M$) via 
\begin{align}
  \nu=&M+2e\sin M+\frac{5}{4}e^2 \sin2M+\frac{e^3}{12}(13\sin3M-3\sin M) \nonumber \\
  &+\frac{e^4}{96}(103\sin4M-44\sin2M), 
\end{align}
which is valid up to the fourth order in eccentricity. Given the eccentricity and mean anomaly for one particle, we remap it onto an eccentric orbit with true anomaly $\nu$. We show the remapping in the horizontal plane schematically in Fig.~\ref{fig:orbits}, where positive eccentricities correspond to upper disc particles ($\zeta>0$) and negative values to lower ones. The fluid column tilts to the left at $\phi=0$ and then to the right at $\phi=\pi$. The velocity on the eccentric orbit is assigned to each particle after remapping.


\begin{figure*}
  \centering
  \includegraphics[width=0.9\linewidth]{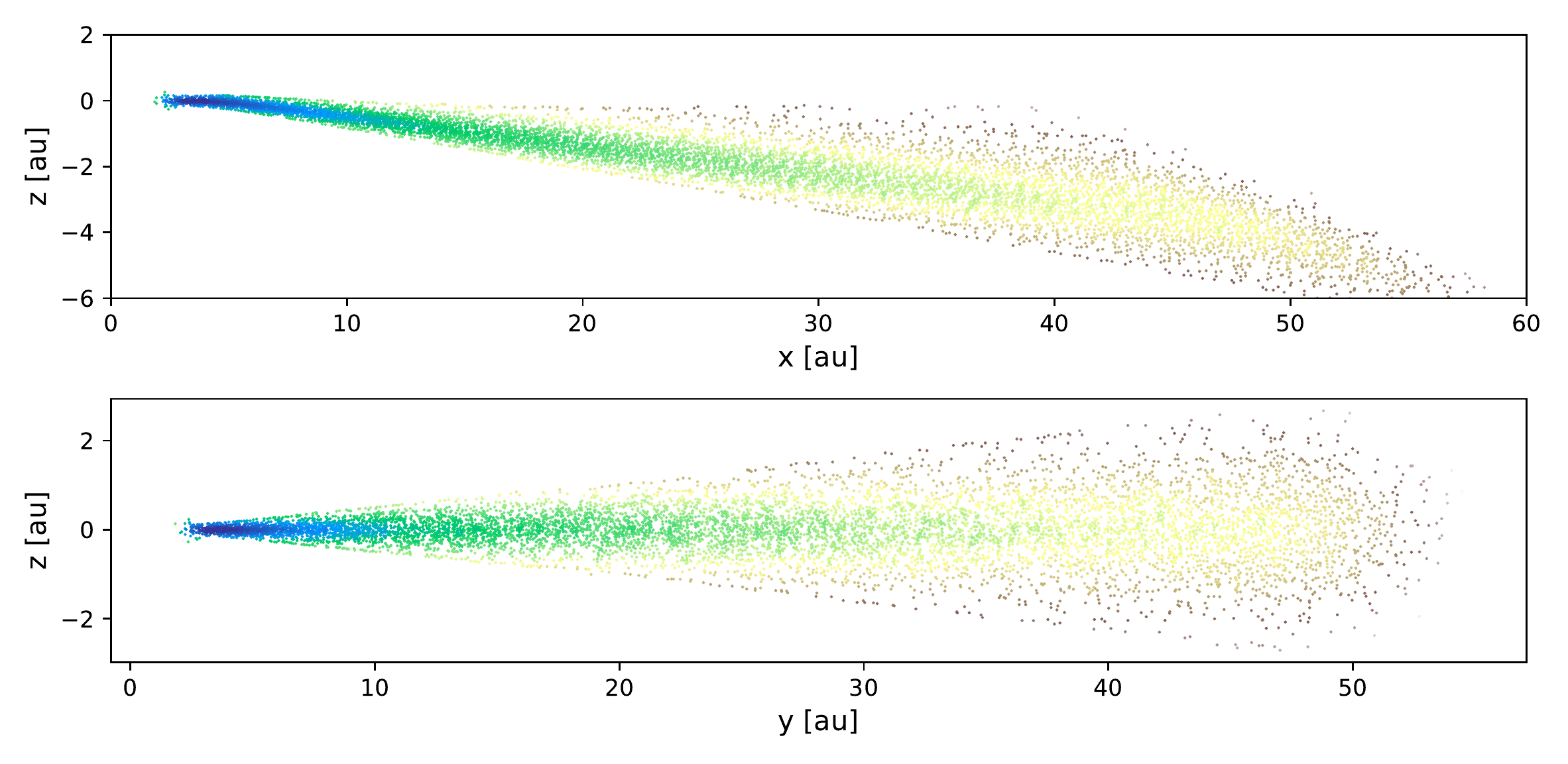}
    \caption{Slice plots (of width $\delta \phi =\pi/500$) showing the particle distribution in the low-amplitude nonlinear warp of Fig.~\ref{fig:NLS} at $\phi=0$ (upper panel) and $\phi=\pi/2$ (lower panel). The colours, from white to blue, indicate the volume density on a logarithmic scale, ranging from $10^{-14}$ to $10^{-10}$ g cm$^{-3}$.}
    \label{fig:rho}
\end{figure*}

Finally, we apply the rotation matrix (obtaining $\beta$ from interpolating NS1 or NS2) to all particles and achieve a representation of the nonlinear bending-wave solution. In Fig.~\ref{fig:rho} we show the particle distribution in vertical slices with $\phi\in(0,\pi/500)$ and $\phi\in(\pi/2,\pi/2+\pi/500)$ for the low-amplitude nonlinear warp. Around $\phi=0$, the upper disc particles with $\zeta, e>0$ are shifted inwards while the lower disc particles are shifted outwards, as shown in Fig.~\ref{fig:orbits}. However, the reference disc is truncated at $\sim 50$ au so that the density field in Fig.~\ref{fig:rho} is slanted at the outer edge. At the edge of the disc, particles are free to diffuse into the vacuum, resulting in a torque-free boundary condition \citep[cf.~the bending-wave reflections at the torque-free inner boundary in][]{Deng2021}. In the outermost part of the disc the internal torque cannot be balanced, forcing the disc tail to precess. However, we focus on the steady-state and torque balance in the disc trunk.

\begin{table}
	\centering
	\caption{Simulations of steady warps around a quadrupole. Here `NS' and `LS' indicate that the initial conditions for the simulations are the nonlinear solutions and the linear solutions discussed in Section~\ref{sec:NLS}. The low- and high-amplitude nonlinear solutions are labelled NS1 and NS2, while the linear solutions are labelled directly by their outer tilt angles. All simulations are adiabatic except the two (nearly) isothermal runs ending with `ISO'.}
	\label{tab:simulations}
	\begin{tabular}{lccr} 
		\hline
		run label & N particles & run time & outer tilt\\
		\hline
		NS1-MFM    & 8M         & 1 ORP     & 0.075\\
		NS1-SPH    & 8M         & 1 ORP     & 0.075 \\
                NS2-MFM    & 8M         & 1 ORP     & 0.13 \\
                NS2-MFM-32 & 32M        & 1 ORP     & 0.13 \\
                NS1-MFM-128&128M       & 0.7 ORPs    & 0.075 \\
                LS0.15-MFM & 8M        & 1.8 ORP    & 0.15\\
                LS0.25-MFM & 8M        & 1.8 ORPs    & 0.25\\
                LS0.25-MFM-ISO & 8M  &1.8 ORPs     &0.25 \\
                LS0.25-SPH & 8M        & 1.8 ORPs    & 0.25\\
                LS0.25-SPH-ISO & 8M        & 1.8 ORPs    & 0.25\\
                LS0.30-MFM & 8M        & 1.8 ORPs    & 0.30\\
                LS0.50-MFM & 8M        & 1 ORP    & 0.50\\
                \hline
	\end{tabular}
\end{table}

\section{Simulations and results}
\label{sec:results}

We construct steady warp models as explained above and evolve the discs for about 1 ORP before the outer discs precess significantly. We run MFM simulations accompanied by SPH simulations for comparison as summarised in Table~\ref{tab:simulations}. The production simulations employ 8M particles, comparable to previous high-resolution SPH simulations \citep{Nealon2015,Drewes2021}.

We focus on the nonlinear warp solutions obtained using the affine model, NS1 and NS2 in Fig.~\ref{fig:NLS}. The low- and high-amplitude nonlinear solutions match to linear warp solutions with an outer tilt 0.1 and 0.3 in the innermost region. Owing to nonlinear effects, the outer tilt is suppressed and equals 0.075 in NS1 and 0.13 in NS2. The highly nonlinear warp NS2 turns out to be challenging for simulations, so we performed a 32M particle simulation to elucidate the effect of resolution in simulations of highly nonlinear warps. The 128M particle simulation explores potential parametric instability in the steady nonlinear warp.

Owing to nonlinear effects, the solution to the linear bending-wave equations is not a true steady state even when the outer tilt is as small as 0.1.  We use the linear bending-wave solution as a form of \emph{arbitrary} initial warp to explore the time-evolution of strong warps. In particular, the affine model predicts a maximum outer tilt of about 0.2, which excludes the existence of very strong warps for the parameters of our reference disc. The ladder of `LS' simulations is intended to explore the boundary of warped disc breaking \citep{Nealon2015,Drewes2021}. Two isothermal simulations are performed to clarify the crucial long-term effects of gas thermodynamics on disc breaking.

\subsection{Low-amplitude nonlinear warp}
\label{sec:low}

\begin{figure}
  \centering
  \includegraphics[width=\linewidth]{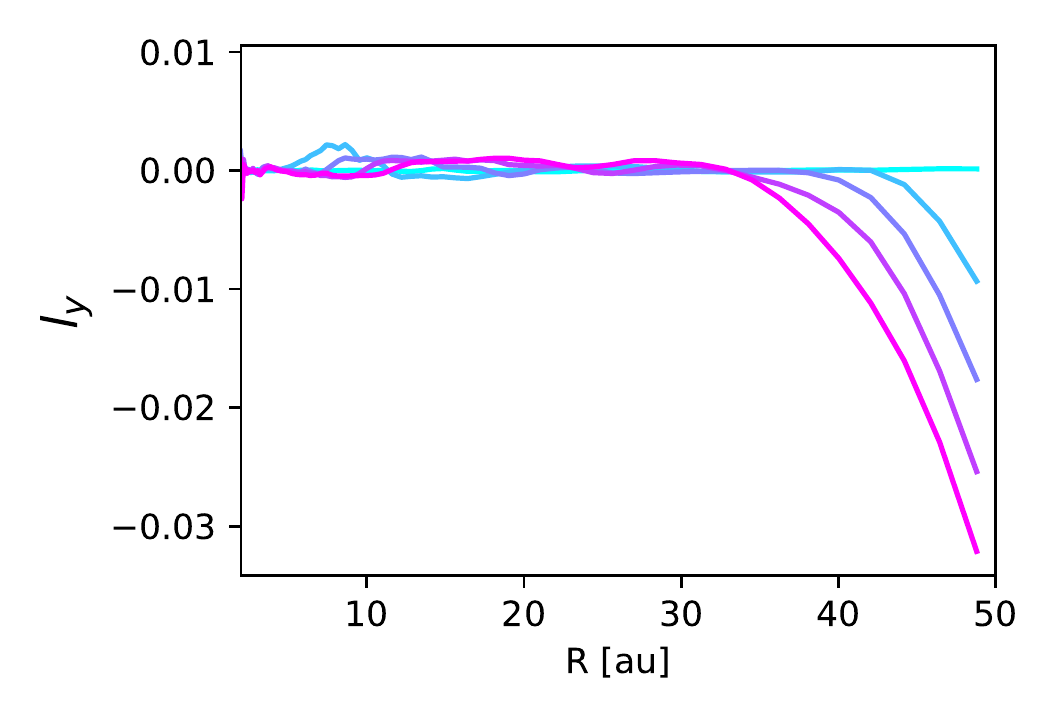}
    \caption{The evolution of $l_y$ every 0.25 ORPs starting from $t=0$ in the NS1-MFM simulation (complementary to Fig.~\ref{fig:lx}). }
  \label{fig:ly}
\end{figure}

Even for the low-amplitude nonlinear solution, the outer tilt is noticeably suppressed compared to the linear solution. Nevertheless, the column deformation is not very significant. In Fig.~\ref{fig:NLS}, the maximum eccentricity at $\zeta=1$ is less than 0.1. The code can handle this state more easily than the other more distorted nonlinear solution. We first check if our 3D simulation maintains a steady warp for this simple case. We focus on the evolution of the tilt vector  $\bm{l}(R,t)$ and the dynamical scale-height $H$~\citep{Ogilvie2018} defined by
\begin{equation}
H^2=\frac{\int \rho (z-Z)^2 \, \mathrm{d}z}{\int \rho \, \mathrm{d}z}.
\end{equation}The  $\bm{l}(R,t)$ profiles are calculated using the specific angular momentum vectors averaged in 100 spherical shells from 2 to 50 au with a logarithmic spacing. The dynamical scale-height is averaged for particles within a vertical slice of angular thickness $\delta \phi=\pi/100$ around $\phi=0$ and $\pi/2$ using the same radial spacing as  $\bm{l}(R,t)$.

We plot the evolution of $l_y$ for NS1-MFM in Fig.~\ref{fig:ly}, which confirms that $l_y$ is very small, except in the outer disc, because the steady warp is untwisted. We observe a small non-zero wave packet of $l_y$ travelling outwards, resulting from the flawed assumption of $\Omega$ being Keplerian in obtaining the steady warp solution. At the outer edge, the disc is truncated (see Fig.~\ref{fig:rho}) and diffuses into the vacuum naturally. The torque-free boundary leads to a torque imbalance in the outer region and thus precession. The evolution of $l_y$ clearly shows the precession of the disc tail on the ORP time-scale, which is about 13 times the rotational period at 9 au.

 \begin{figure*}
  \centering
  \includegraphics[width=0.95\linewidth]{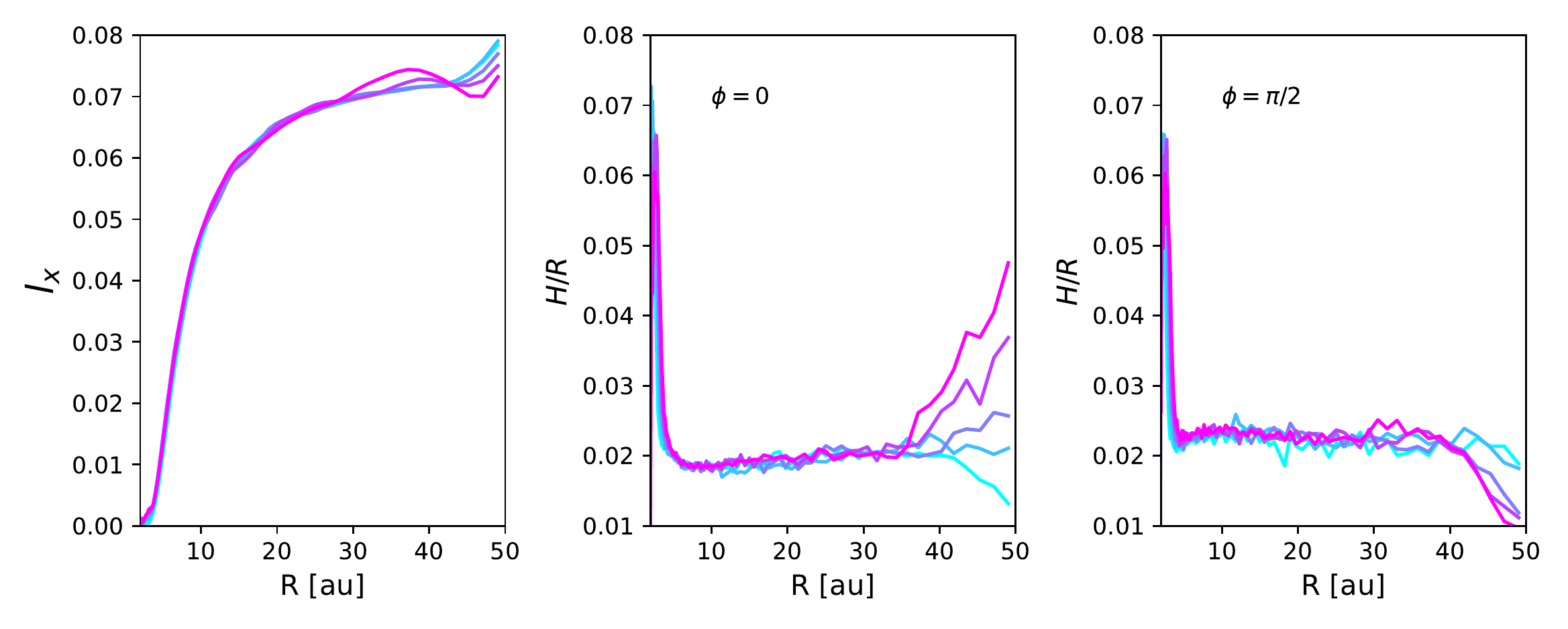}
    \caption{The evolution of $l_x$ and the dynamical scale-height at $\phi=0$ and $\phi=\pi/2$ are shown for the NS1-MFM simulation. The lines show the profiles every 0.25 ORPs, starting from $t=0$, with warm colors corresponding to later times.}  
  \label{fig:lx}
\end{figure*}

\begin{figure*}
  \centering
  \includegraphics[width=0.9\linewidth]{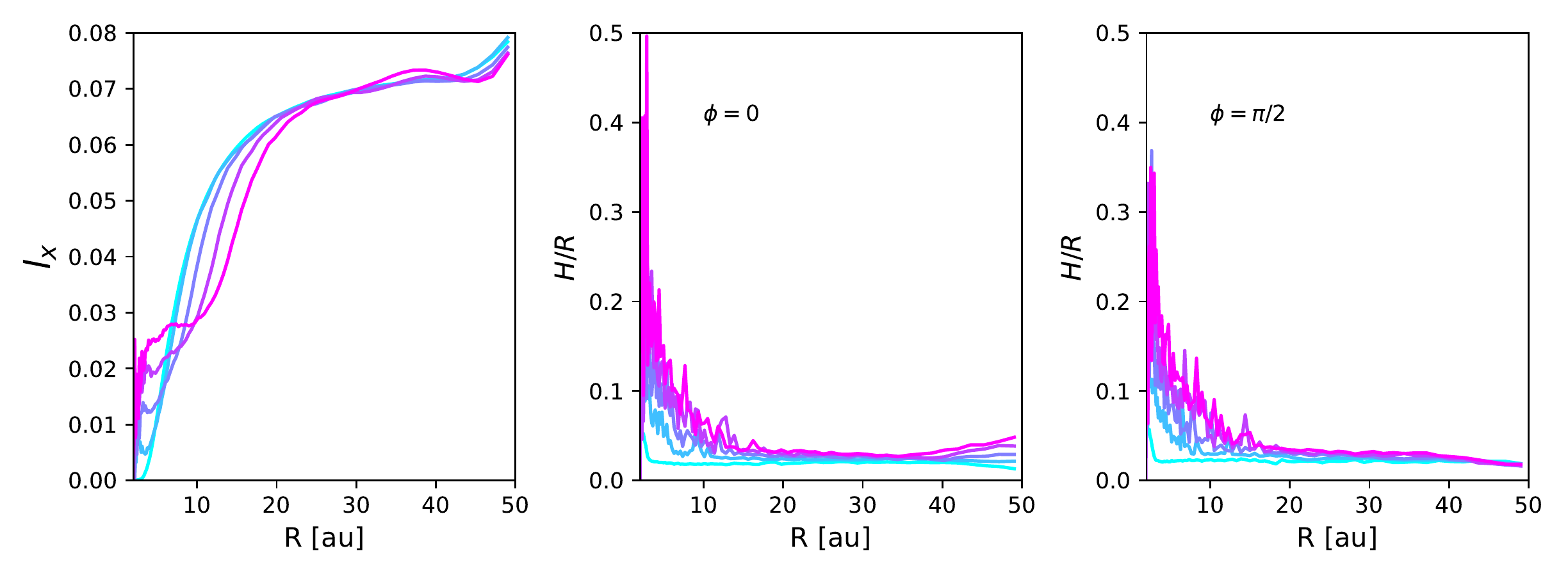}
  \caption{Similar to Fig.~\ref{fig:lx}, but for the NS1-SPH simulation. Both $l_x$ and $H$ evolve significantly, in contrast to the NS1-MFM simulation.}
  \label{fig:lxsph}
\end{figure*}

That said, in Fig.~\ref{fig:lx}, $l_x$ evolves little and the warp profile is well maintained in the disc trunk, barely affected by the precession of the disc tail. The model is also in vertical hydrodynamic equilibrium as the dynamical scale-height remains unchanged in the disc trunk. Note that the disc is compressed or expanded along the $\phi=0$ and $\phi=\pi/2$ axes so that $H/R$ is smaller or larger than $\epsilon=0.02$, respectively.


A similar simulation using SPH, NS1-SPH, cannot maintain the nonlinear steady warp. In Fig.~\ref{fig:lxsph}, the warp profile is reasonably well preserved in the first 0.25 ORP, indicating a correct initialization of the nonlinear warp. However, $l_x$ evolves significantly after 1 ORP in the $R<20$ au region although the outermost region shows only moderate precession similar to NS1-MFM. In short, SPH can only sustain the steady nonlinear warp briefly before it is destroyed somehow.  The expected vertical equilibrium is also disrupted, and we observe a consistent disc expansion in the vertical direction. The disc thickening is caused by accumulative heating from numerical dissipation in the NS1-SPH simulation. 

\begin{figure}
  \centering
  \includegraphics[width=\linewidth]{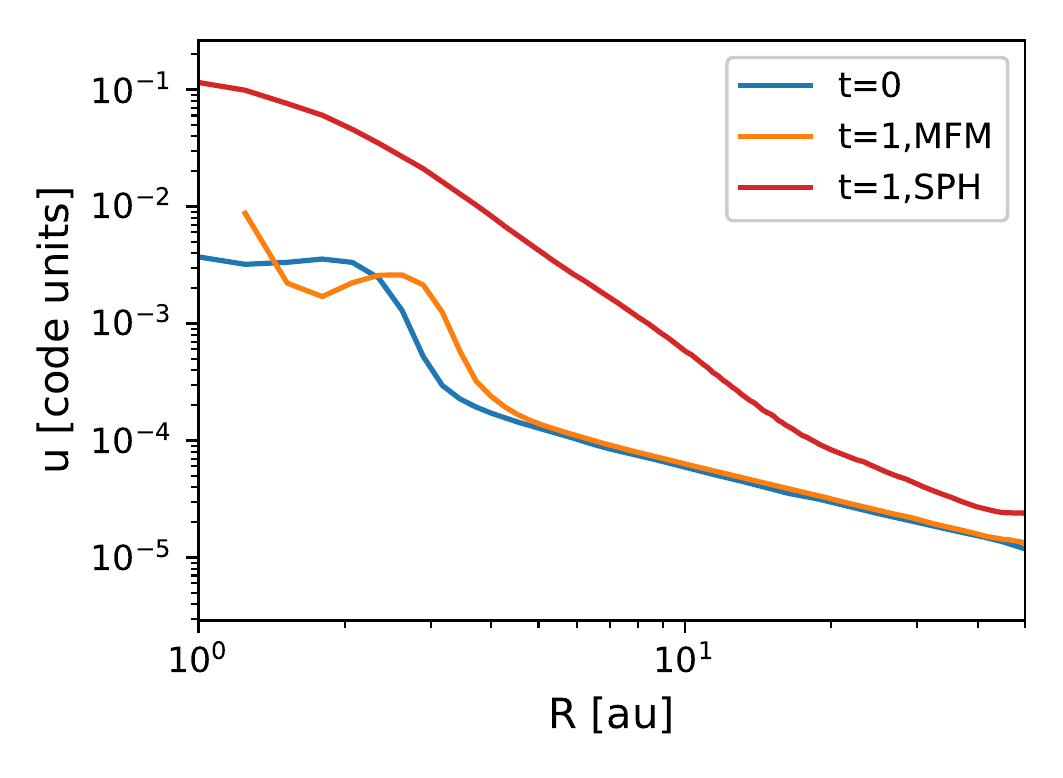}
  \caption{The internal energy profile at the end of the simulations (1 ORP) for the low-resolution NS1 model. In the NS1-SPH simulation, the disc is significantly heated so that the disc also becomes thicker in Fig.~\ref{fig:lxsph}.}
  \label{fig:NS1u}
\end{figure}
\begin{figure*}
  \centering
  \includegraphics[width=\linewidth]{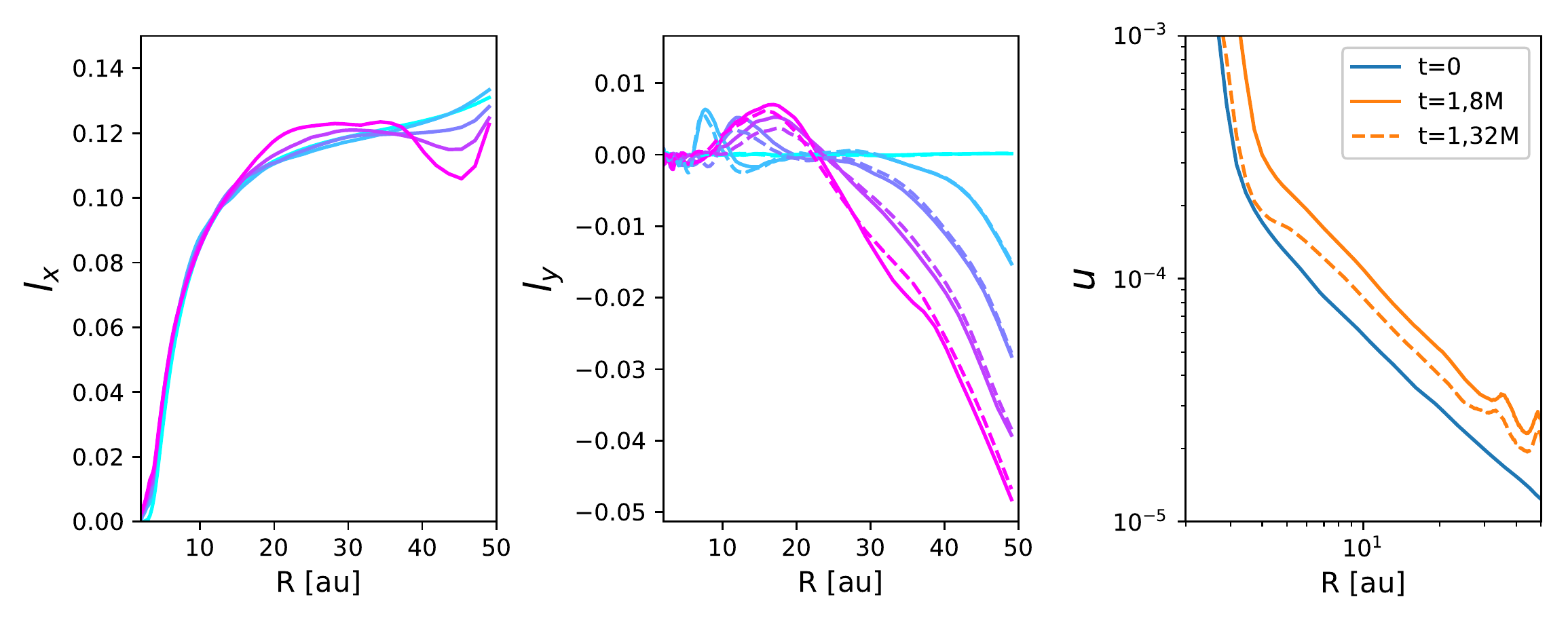}
  \caption{The tilt vector evolution (left and middle panels) in the high-amplitude nonlinear warp simulations, NS2-MFM (solid lines) and NS2-MFM-32 (dashed lines). For clarity, we omitted the $l_x$ profiles of NS2-MFM-32 in the left panel. The lines with warmer colors are for later times starting from $t=0$ every 0.25 ORPs. The internal energy at the end of the simulations (1 ORP) is compared with the initial value in the right panel.}
  \label{fig:amp03}
\end{figure*}

In Fig.~\ref{fig:NS1u} we plot the internal energy change after 1 ORP in the MFM and SPH simulation of NS1. The internal energy increases substantially throughout the disc in the SPH simulation. It also forms a steeper internal energy profile because the inner disc executes more rotations and thus deposits more numerical heating than the outer disc. In contrast, the internal energy profile in the NS1-MFM simulation is almost unchanged except near the poorly resolved inner edge. We attribute this excessive numerical heating to artificial viscosity in our SPH implementation, and discuss resolution effects and numerical dissipation later in Section~\ref{sec:col}.

\subsection{High-amplitude nonlinear warp}

The higher-amplitude nonlinear solution (NS2) has larger eccentricities and more substantial deformations compared to NS1. As shown in Fig.~\ref{fig:H03}, the disc can be either compressed or expanded significantly in the vertical direction in addition to the eccentric orbital motion. The complex flow poses significant challenges to simulations. Notably, the low-density disc surface region with $\zeta>3$ is less well resolved than the mid-plane region in any Lagrangian method. Increasing the total number of particles leads to only a minor improvement in resolution in surface region. As a result, modelling the highly eccentric streamlines (see Fig.~\ref{fig:NLS}) at high altitudes is a likely bottleneck in Lagrangian simulations. Owing to the poor performance of NS1-SPH, we avoided SPH simulations for NS2 and instead carried out an extra MFM simulation with 32M particles to study the effects of resolution  (see Table~\ref{tab:simulations}).

We show the evolution of the tilt vector and internal energy in Fig.~\ref{fig:amp03}. The $l_x$ profile is less well preserved than in the low-amplitude warp simulation in Fig.~\ref{fig:lx}. The deviation is modest for $R<20$ au but more significant beyond 20 au, which we attribute to boundary effects and precession due to the torque-free outer edge. 
The NS2-MFM model has high eccentricities for particles near the outer edge, so the density field is more slanted than the NS1-MFM model at $\phi=0$ (see Fig.~\ref{fig:rho}). For example, if one particle lies at 50 au with $\zeta=3$ in the reference state it will be mapped to about 25 au in the NS2 model (see Figs \ref{fig:NLS} and \ref{fig:orbits}). As we know for NS1-MFM, the slanted disc tail with non-zero net torque precesses on the ORP time-scale. In Fig.~\ref{fig:amp03}, the $l_y$ evolution further confirms precession in a large region ($R>20$ au) owing to the torque imbalance.

In Fig.~\ref{fig:amp03}, the disc becomes hotter after 1 ORP in the NS2-MFM simulation than it does in NS1-MFM in Fig.~\ref{fig:NS1u}, presumably because of greater numerical dissipation in the high-amplitude warp. Increasing the number of particles to 32M diminishes the numerical heating to some extent. However, there is little difference in the warp evolution on quadrupling the number of particles, judging by the development of $l_y$ in Fig.~\ref{fig:amp03}. The high-amplitude nonlinear warp with complex flow challenges direct numerical simulations. Our MFM simulation can reasonably simulate it while significant numerical heating is observed. We can further increase the resolution to beat down the numerical dissipation, but the parametric instability will kick in at sufficiently high resolution and muddy the waters (see Section~\ref{sec:PI}).

\subsection{Evolution of fluid columns}
\label{sec:col}

\begin{figure*}
  \centering
  \includegraphics[width=\linewidth]{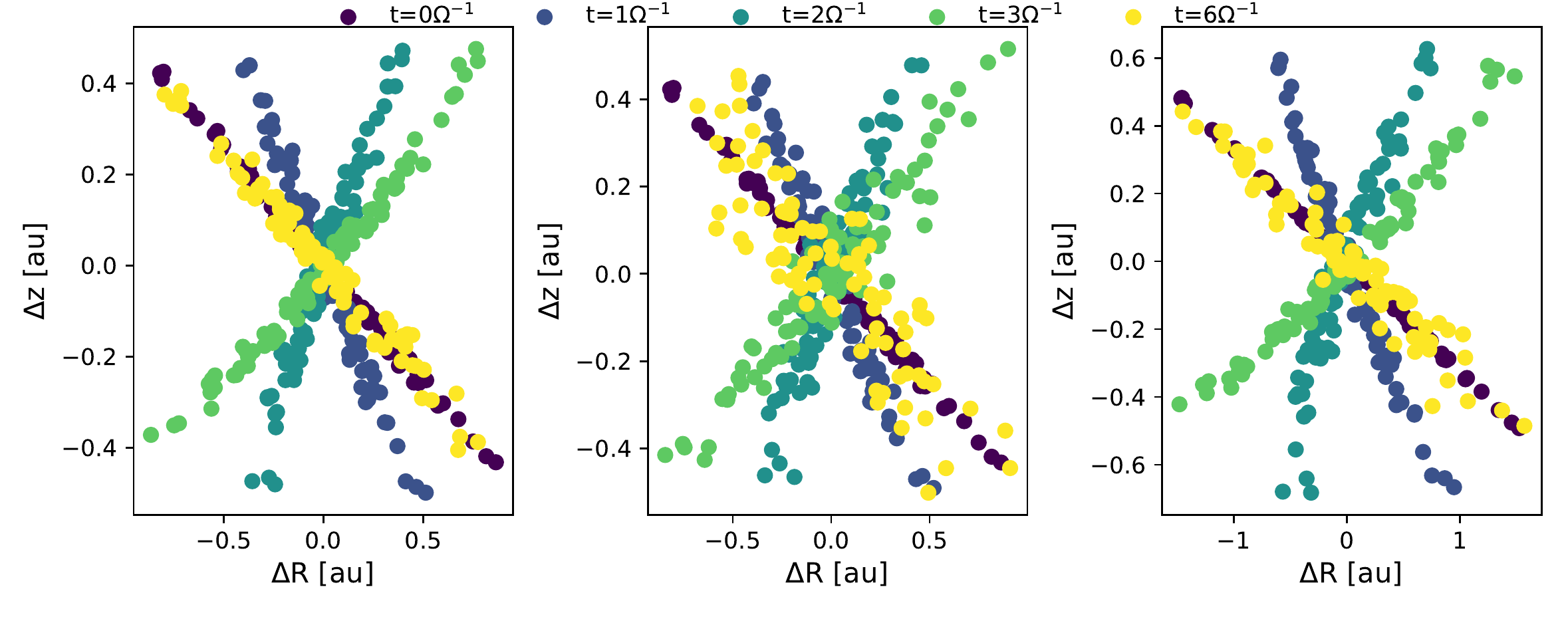}
  \caption{The evolution of fluid columns in the co-moving frame (at 9 au) in the NS1-MFM, NS1-SPH and NS2-MFM simulations, from left to right (see text). The columns start at $\phi=0$ and the time is given in units of the local dynamical time $\Omega^{-1}$.}
  \label{fig:col}
\end{figure*}

The affine model describes the motion of a disc as the translations and linear transformations of its fluid columns. To illustrate the affine model directly,  we visualize the evolution of an exemplary fluid column here, taking advantage of our Lagrangian methods. We choose a fluid column in the initial reference state (Fig.~\ref{fig:IC}) spanning 8.9--9.0 au radially and 0--$\pi/100$ in the azimuthal direction. The fluid column is sampled by 81, 327 and 1273 particles in the 8M, 32M, and 128M particle reference models, respectively. We follow the trajectories of all particles in this column but focus on their distribution and averaged statistics. We show in Fig.~\ref{fig:col} the motion of the fluid column, in a frame moving with the centre of the column, for NS1-MFM, NS1-SPH and NS2-MFM.

The fluid column is tilted anti-clockwise at $\phi=0$ (see Fig.~\ref{fig:orbits}). The column in the NS2-MFM simulation is tilted more significantly owing to larger eccentricities at a given $\zeta$ than in NS1-MFM. As the column orbits around the centre of the potential, it tilts back and forth and experiences vertical expansion and compression periodically. After one full rotation ($\phi=2\pi$), the column should restore its shape. Indeed, in the NS1-MFM simulation, the column nearly repeats itself at $t=6\Omega^{-1}$. In contrast, the particles in the NS1-SPH simulation are significantly scattered and diffuse perpendicularly to the column. Even the strongly nonlinear NS2-MFM model preserves the fluid column better than the NS1-SPH simulation.

\begin{figure}
  \centering
  \includegraphics[width=\linewidth]{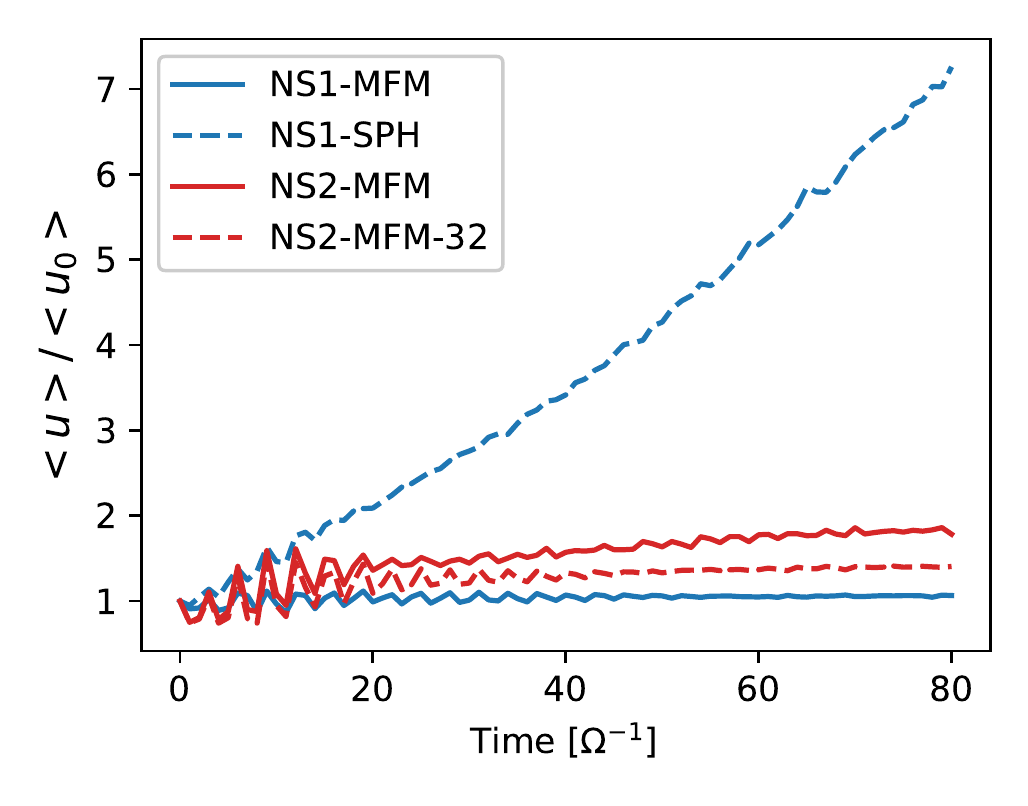}
  \caption{The evolution of the averaged internal energy (normalised to the initial value) in the fluid column, spanning 8.9--9.0 au and $\phi\in(0,\pi/100)$ in the unwarped reference state, for several models in table \ref{tab:simulations}.  The time is given in units of the local dynamical time $\Omega^{-1}$ at 9 au.}
  \label{fig:colu}
\end{figure}

We present the averaged specific internal energy for particles within the aforementioned column in Fig.~\ref{fig:colu} for different simulations. In the NS1-MFM model, the internal energy oscillates because of the expansion and compression of the column and eventually settles down to a slightly larger value. The higher-amplitude warp in NS2 leads to more substantial expansion and compression and thus stronger internal energy fluctuations. The column tilting and deformation (see Fig.~\ref{fig:col}) can lead to irreversible numerical heating. The better-resolved NS2-MFM-32 model shows less numerical heating than NS2-MFM, but it still cannot keep the internal energy structure. The numerical dissipation is devastating in the SPH simulation and the internal energy increases rapidly. It is no wonder that the steady warp corrupts quickly in the NS1-SPH simulation. If an isothermal equation of state were adopted in SPH simulations, the internal energy errors would not accumulate as in Fig.~\ref{fig:colu}, but the oscillatory flow would be damped similarly. 

The complex flow in the inviscid warped disc poses a substantial challenge for numerical modelling. The rapid numerical dissipation in our SPH simulation eventually leads to the degradation of the steady state. There have been continuing efforts to construct \emph{inviscid} SPH algorithms in which the artificial viscosity is minimal away from shocks \citep{Balsara1995, Cullen2010}. The SPH method in GIZMO adopts a modified version of the Cullen \& Dehnen switch; it seems that the switch wrongly identifies the oscillatory flow in a warped disc as a shock during the compressional phase and applies an excessive artificial viscosity (see Fig.~\ref{fig:colu}). In our simulation, we set a floor for the artificial viscosity $\alpha_\text{SPH}$ parameter being 0.05 \citep{Hopkins2015}, which translates to a non-zero physical viscosity \citep[see equation 6 in][]{Meru2012}, depending on resolution. It is debated whether a floor for $\alpha_\text{SPH}$ is beneficial in SPH simulations~\citep{Cullen2010,Rosswog2015,Wadsley2017,Price2018}. The most appropriate choice of the $\alpha_\text{SPH}$ parameter probably also depends on the physical application. We followed previous studies of inviscid warps with SPH to maintain a floor value for $\alpha_\text{SPH}$ \citep[see, e.g.,][]{Facchini2013,Drewes2021}. We note that recent high-resolution (100M particles) SPH simulations managed to achieve a physical $\alpha$ between 0.002 and 0.02 for a warped disc around a black hole \citep[see figure 2 of][]{Drewes2021}.

The finite-difference method employing artificial viscosity also found excessive dissipation in simulations of strong warps \citep{Sorathia2013}.  Godunov-type grid codes using a finite-volume scheme, free of artificial viscosity, have been deployed on warped disc simulations \citep[see, e.g.,][]{Fragner2010,Liska2019, Dyda2020}. However, advection errors can be severe when the grid is not aligned with the flow \citep{Hopkins2015}. The eccentric and oscillating flow in a warped disc does not align well even with spherical coordinates. Therefore dedicated simulations are warranted to assess how well different grid codes capture the oscillatory shearing flow in a nonlinear warped disc.

\subsection{Arbitrary warps}
\label{sec:break}

\begin{figure}
  \centering
  \includegraphics[width=\linewidth]{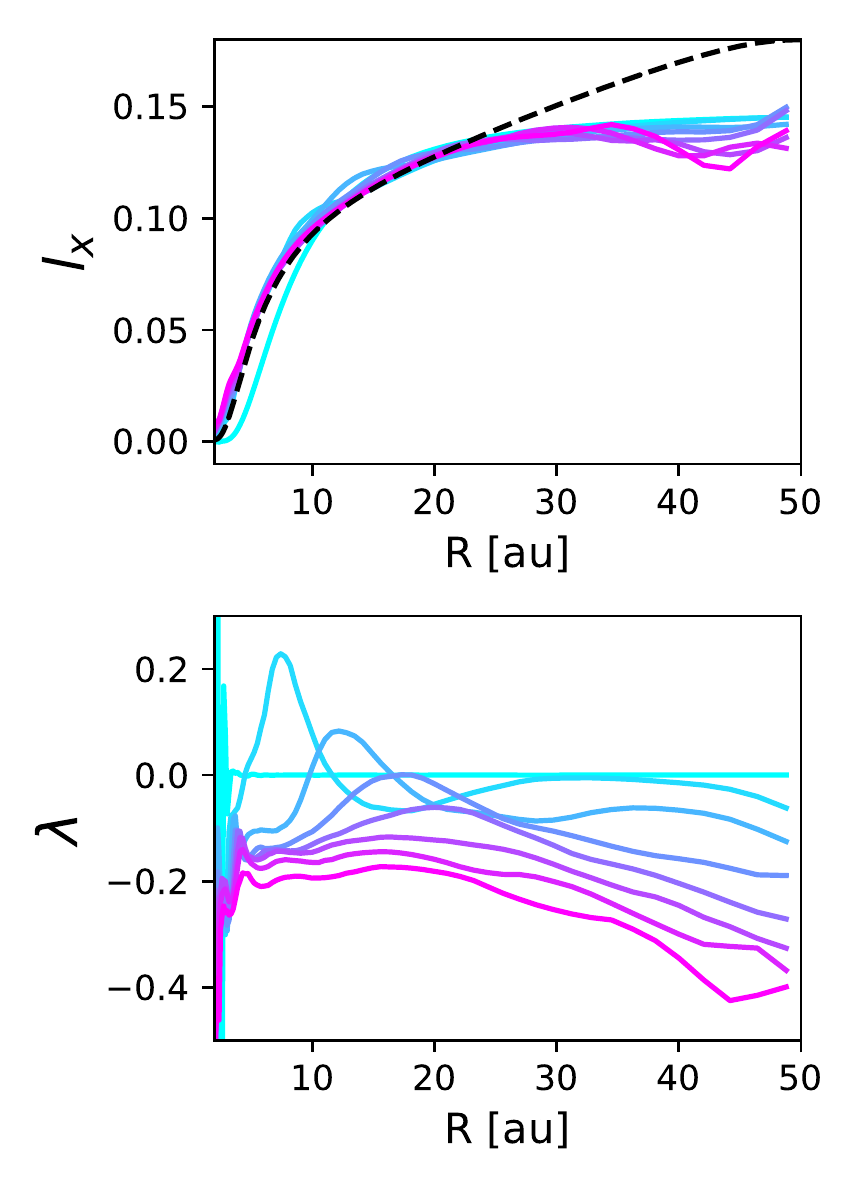}
  \caption{The evolution of $l_x$ (similar to Fig.~\ref{fig:lx}) and the precession angle $\lambda$ (in radians) for LS0.15-MFM, every 0.25 ORPs between $t=0$ and $t=1.75$ ORPs; warmer colors for later times. The black dashed line denotes a nonlinear steadily precessing steady warp.}
  \label{fig:lx15}
\end{figure}

We run \emph{steady} disc models predicted by the linear inviscid warp theory with various outer tilts (see equation \ref{eq:LS}) as summarized in Table~\ref{tab:simulations}. However, as we know, even the linear warp with an outer tilt of 0.1 is quite far from a true steady state (see Fig.~\ref{fig:NLS}). Therefore, the linear warp profile should not be considered a steady solution but rather a form of arbitrary initial warp.

\subsubsection{Low-amplitude arbitrary warps}

Low-amplitude warps, regardless of the initial profiles, evolve towards steady solutions to the 1D linear bending-wave equations around a black hole \citep{Lubow2002,Drewes2021} or binary \citep{Facchini2013}. This tendency of linear bending waves is confirmed by 3D hydrodynamic simulations. For example, the steady-state oscillatory tilt around a black hole gradually establishes itself starting from a uniform tilt in 3D simulations \citep{Nealon2015, Drewes2021}. Nevertheless, how nonlinearity alters this picture is unclear owing to the lack of nonlinear steady solutions for inviscid warps.

\begin{figure}
  \centering
  \includegraphics[width=\linewidth]{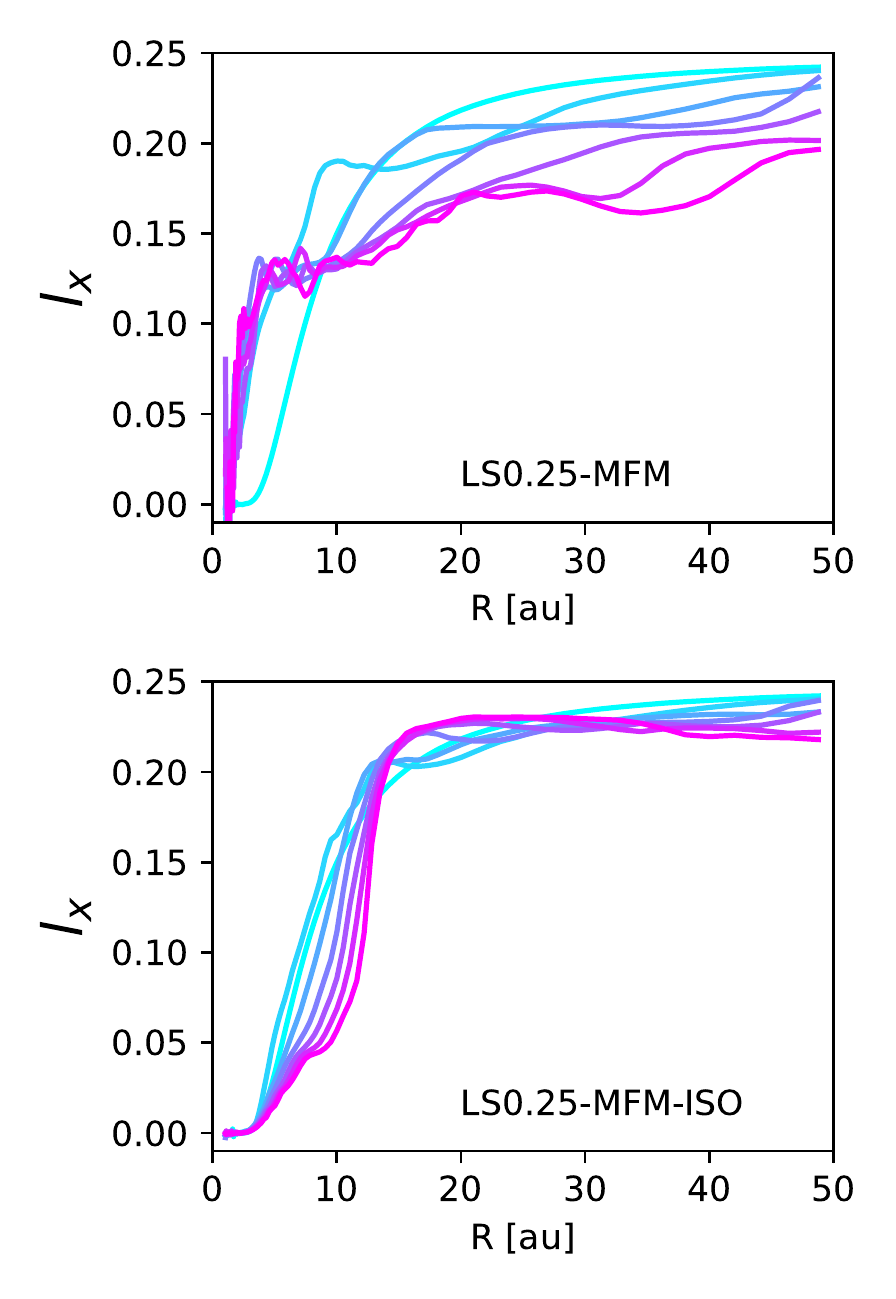}
  \caption{The evolution of $l_x$ for LS0.25 models every 0.25 ORPs between $t=0$ and $t=1.75$ ORPs; warmer colors for later times. The thermodynamics plays a crucial role on the long-term disc evolution.}
  \label{fig:lx25}
\end{figure}

In the low-amplitude model, LS0.15-MFM, the disc quickly evolves away from the initial warp profile as shown in Fig.~\ref{fig:lx15}. Later, a quasi-steady $l_x$ profile is gradually established in the inner disc ($R<20$ au). In Fig.~\ref{fig:lx15}, the disc precession angle is calculated from $\lambda=\text{arctan}(l_y/l_x)$. Differential precession occurs around 10 au in the early stages but the inner disc precesses uniformly near the end of the simulation. This transition is violent. The disc is significantly heated by shocks between colliding fluid columns although the numerical heating is hard to assess during this violent rearrangement. At the end of the simulation, the disc possesses a specific internal energy profile close to $u=0.004\,R^{-1.4}$ in the disc trunk, corresponding to $H_0/R=0.052\,R^{-0.2}$.

\begin{figure*}
  \centering
  \includegraphics[width=\linewidth]{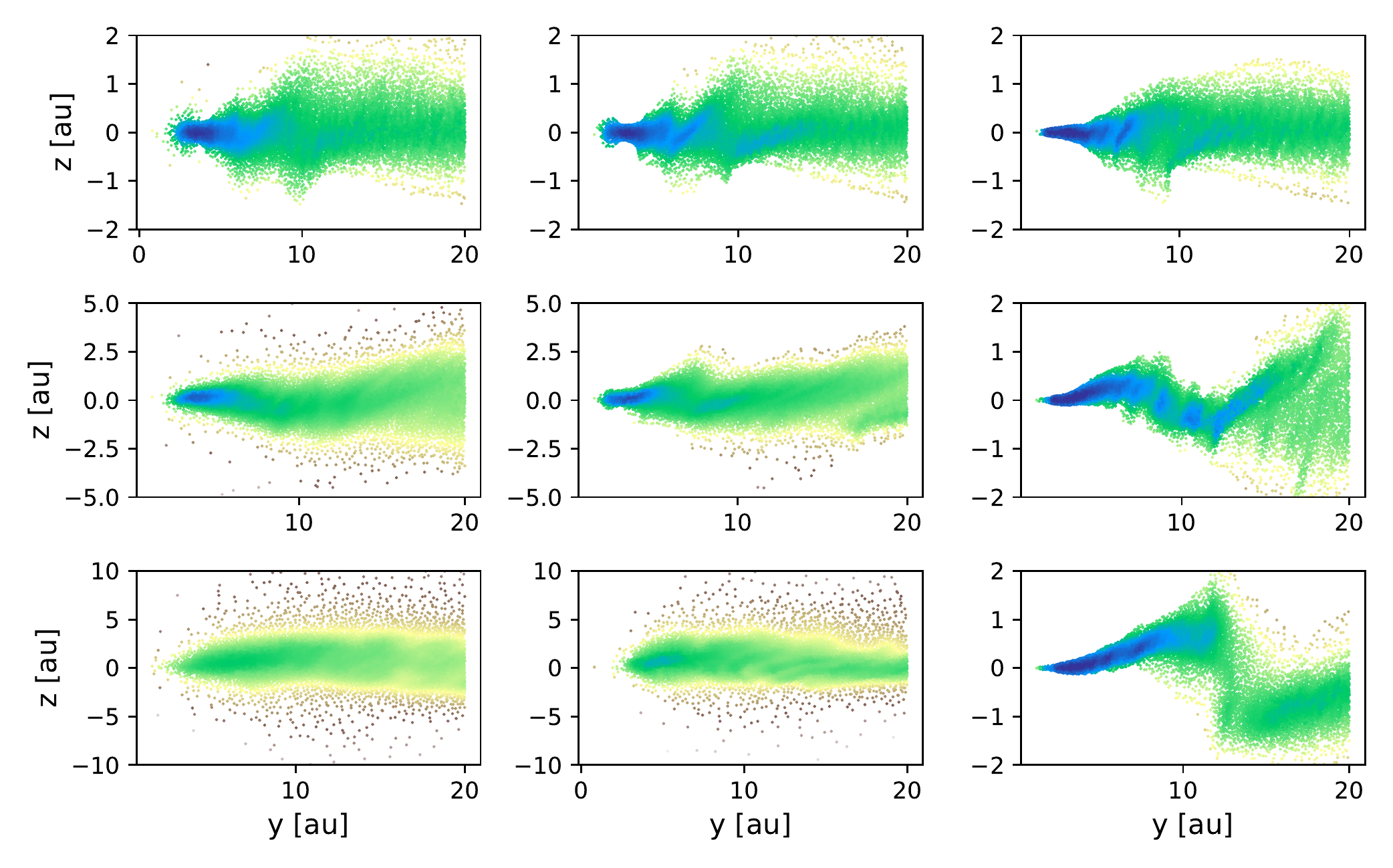}
  \caption{The distribution of particles with $\phi \in (\pi/2,\pi/2+\pi/100)$ in the LS0.25 simulations using (from left to right) SPH, MFM and MFM-ISO, at different times. The top, middle and bottom rows show snapshots at times of $5$, $20$ and $120$, respectively, in units of the dynamical time $\Omega^{-1}$ at 9 au. The colors, from white to blue, indicate the volume density on a logarithmic scale from $10^{-15}$ to $10^{-10}$ g cm$^{-3}$. The isothermal simulation is markedly different from its adiabatic equivalents.}
  \label{fig:break}
\end{figure*}

We solve for nonlinear steady warps around the quadrupole with the updated internal energy profile. Here we consider a finite disc whose density is tapered to zero at 2 au and 50 au with $\tanh$ functions of width $H_0$ to better match the simulation. The closest equivalent to the steady warp considered for the finite disc is a rigidly precessing disc that is steady in a slowly rotating frame. We compute the solution by generalizing equations (\ref{ode1}) and (\ref{ode2}) to include the global precession rate $\omega$, which is regarded as an eigenvalue.\footnote{This modification amounts to adding a term $(\omega/\Omega)\Sigma_0(1-e_1^2)^{1/2}\sin\beta$ to the right-hand side of equation~(\ref{ode1}) and a term $-(\omega/\Omega)r^2e_1(1-e_1^2)^{-1/2}\cos\beta$ to the right-hand side of equation~(\ref{ode2}). The value of $\omega$ is determined by a shooting method that forces the solution to be regular at the torque-free outer boundary where $\Sigma_0$ vanishes.} We find that a solution with an outer tilt of 0.18 radians can reproduce the $l_x$ profile within 20 au. The predicted (retrograde) global precession rate is $-\omega=1.9\times 10^{-4}\Omega(1)$, while we measure an averaged precession rate of $6.2\times 10^{-5}\Omega(1)$ between $t=1.5$ and 1.75 ORPs for the inner disc. The discrepancy is probably caused by the short simulation time so that a uniform precession is not established for the $R>20$ region yet (see Fig.~\ref{fig:lx15}). Indeed we observe an increase in the inner disc's precession rate with time. This evidence suggests that a low-amplitude arbitrary initial warp does evolve towards a nonlinear steady (or steadily precessing) warp.


\subsubsection{High-amplitude arbitrary warps}

We focus on models with an initial outer tilt greater than the maximum value of about 0.2 predicted by the affine model. The evolution of these initial warps may shed light on the breaking of low-viscosity warped discs \citep{Fragner2010} since they cannot achieve a nonlinear steady-state. Disc tearing is common in simulations of strongly nonlinear warps in the diffusive regime \citep[see, e.g.][]{Lodato2010,Nixon2012} and is thought to be due to a viscous-warp instability \citep{Ogilvie2000, Dogan2018}. The breaking of low-viscosity warped discs is less well understood but similar disc breaking occurs for an outer misalignment $>30^{\circ}$ \citep{Nealon2015,Drewes2021} around a spinning black hole or $>40^{\circ}$ around a binary \citep{Facchini2013}.

The three high-amplitude simulations, LS0.25-MFM, LS0.30-MFM, and LS0.5-MFM, have a similar evolutionary track, which is apparently different from LS0.15-MFM. Therefore, we take LS0.25-MFM and its variants in Table~\ref{tab:simulations} as an example.  We plot the $l_x$ evolution for LS0.25-MFM and LS0.25-MFM-ISO in Fig.~\ref{fig:lx25}. In the adiabatic simulation, $l_x$ in the inner disc steepens with time, while in the outer disc it evolves towards a flat tilt. We note that the two higher-amplitude models, LS0.30-MFM and LS0.50-MFM, show a similar but faster evolution. The isothermal disc shows a different evolution compared to the adiabatic simulations represented by LS0.25-MFM. In Fig.~\ref{fig:lx25}, its $l_x$ profile steepens into a step function similar to previous isothermal simulations of warped disc breaking~\citep{Nealon2015,Drewes2021}. It is noteworthy that more substantial tilt is transported to the inner disc in the adiabatic simulation than its isothermal counterpart.

\begin{figure*}
  \centering
  \includegraphics[width=\linewidth]{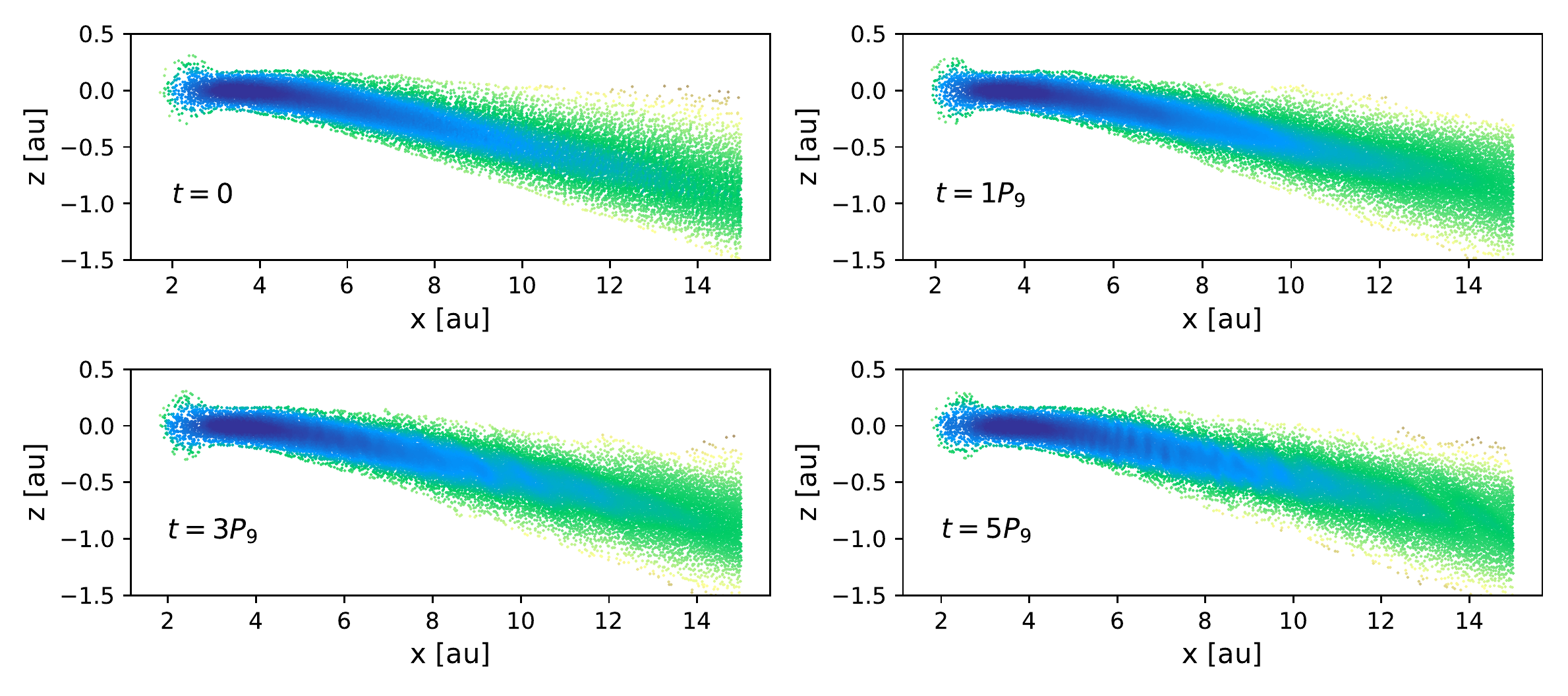}
  \caption{The growth of parametric instability in the NS1-MFM-128 simulation, as indicated by oblique stripes in the density field [slice plot with $\phi \in (0, \pi/1000)$].  The colours, from white to blue, indicate the volume density on a logarithmic scale from $10^{-14}$ to $10^{-10}$ g/cm$^{-3}$. The time unit $P_9$ is the local rotational period at 9 au.}
  \label{fig:PI}
\end{figure*}

To reveal the difference in adiabatic and isothermal simulations, we plot the  flow structure in Fig.~\ref{fig:break}. The LS0.25 warp drives strong horizontal flows that tear up the disc as illustrated in Fig.~\ref{fig:break}. On the local dynamical time-scale, the disc breaks into connected parts along the $y$-axis owing to differential precession (see lower panel of Fig.~\ref{fig:lx15}). They interact actively and heat the disc significantly in the adiabatic simulations. In the long term, the broken pieces reconnect smoothly in the adiabatic simulations. Similar processes are observed in other higher-amplitude MFM models on a shorter time-scale. LS0.25-SPH, the SPH model for comparison, shows qualitatively similar behaviour but less sharp density jumps in Fig.~\ref{fig:break}. The disc is also substantially hotter than the MFM simulation at the same stage, owing to greater numerical dissipation (see Fig.~\ref{fig:colu}). The reconnection of broken discs in our adiabatic simulations is distinct from previous, isothermal studies.

The isothermal simulation, LS0.25-MFM-ISO shows similar disc breaking at early times but the inner and outer disc become more isolated with time (see Fig.~\ref{fig:break}). This scenario resembles previous disc tearing simulations in the bending-wave regime \citep{Nealon2015,Facchini2013,Drewes2021}. However, we caution that some early studies may have excessive numerical viscosity leading to diffusive warp evolution \citep[see section 2.4.1 in][]{Drewes2021}. Indeed, \cite{Nealon2015} showed (their figure 8) that low-resolution (diffusive) SPH simulations may prevent disc breaking. We confirmed this by running LS0.25-SPH-ISO, which is similar to LS0.25-MFM-ISO but does not break into distinct rings.

We report a critical disc breaking angle as low as $14^{\circ}$, primarily due to the disc's thinness (which increases the nonlinearity parameter~$\chi$) and partly due to the relatively low numerical viscosity in MFM (see Section \ref{sec:col}). We note that the affine model predicts a larger maximum tilt of $20^\circ$ for a thicker disc with $\epsilon=0.05$. The breaking radius found in simulations roughly coincides with the most nonlinear radius with the largest $\chi$. On the other hand, thermodynamics plays a decisive role on the long-term disc breaking but even our MFM simulations probably have exaggerated heating in this violent process (see Section \ref{sec:col}). High-resolution simulations with more realistic thermodynamics are crucial for future warped disc breaking studies.

\subsection{Parametric instability}
\label{sec:PI}

\begin{figure}
  \centering
  \includegraphics[width=\columnwidth]{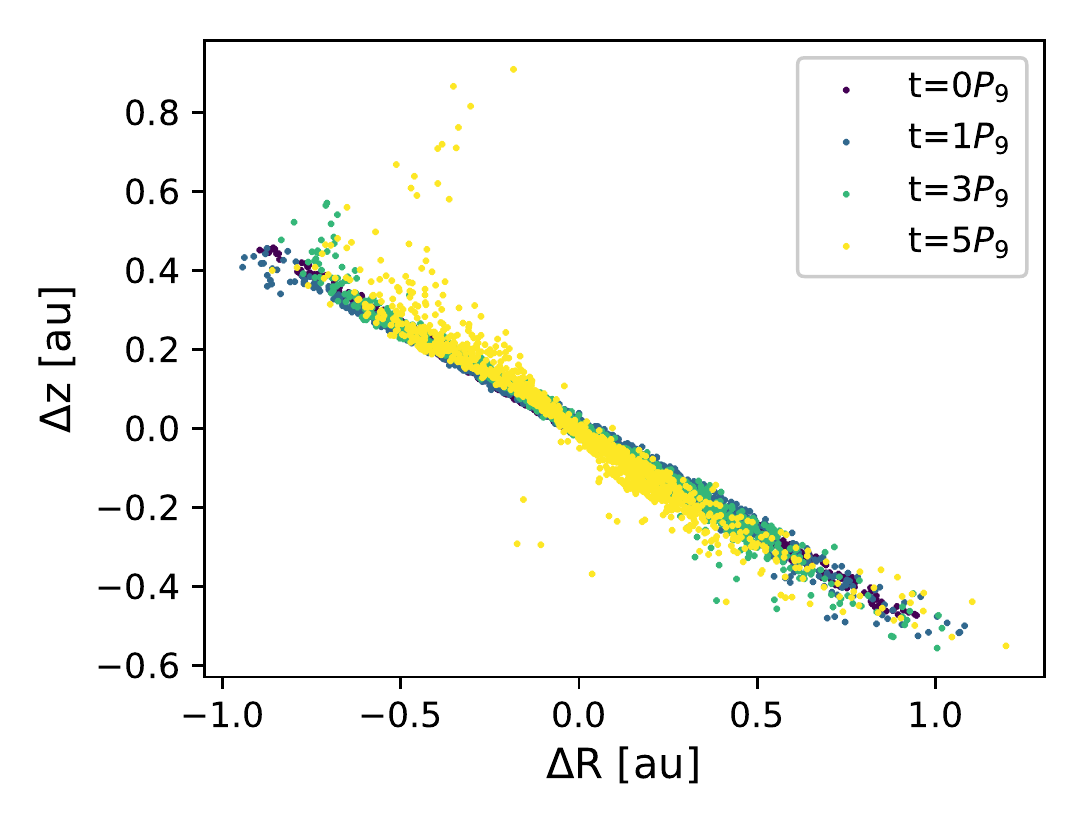}
  \caption{The evolution of a fluid column in the co-moving frame (at 9 au) in the NS1-MFM-128 simulation. The column is disrupted by the parametric instability at later times.}
  \label{fig:col128}
\end{figure}

The parametric instability is expected in our steady nonlinear warp models \citep{Gammie2000, Ogilvie2013b}. \cite{Deng2021} showed that MFM can marginally capture parametric instability at a resolution comparable to NS1-MFM-128. Indeed, in Fig.~\ref{fig:PI}, we observe a secondary instability causing characteristic oblique density structures resembling those in previous parametric instability simulations \citep{Gammie2000,Paardekooper2019,Deng2021}. In Fig.~\ref{fig:col128}, we further analyse the evolution of a fluid column initially spanning 8.9--9.0 au and 0--$\pi/100$ in the reference state (see Section~\ref{sec:col}). This column is similar to the one in Fig.~\ref{fig:col} for NS1-MFM but is resolved with about 16 times more particles. The fluid column repeats itself almost perfectly in the first rotation (see also Fig.~\ref{fig:col}). After three local orbits, the flow becomes unstable and particles are scattered out of the fluid column. 

\begin{figure}
  \centering
  \includegraphics[width=\columnwidth]{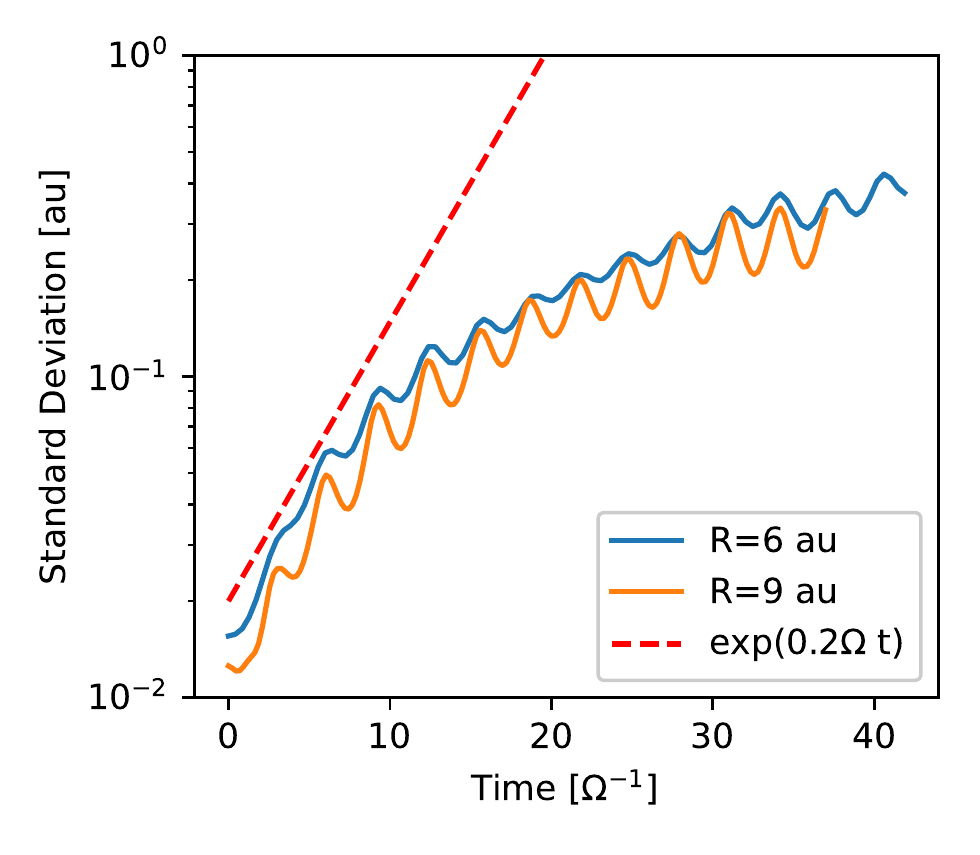}
  \caption{The evolution of the standard deviation of particle positions in fluid columns centred at 6 au and 9 au. Here $\Omega^{-1}$ is the local dynamical time-scale at either 6 au or 9 au. The dashed line shows an approximate fit to the initial exponential growth.}
  \label{fig:std}
\end{figure}

We measure the particle displacement from the fluid column to check the growth rate of the parametric instability. We first fit a straight line to the particle positions in the frame moving with the centre of the column. This is done by fitting the $x$ and $y$ coordinates to the $z$ coordinates with a least-squares regression. The standard error of the fitting is thus a measure of the particle dispersion. In Fig.~\ref{fig:std}, we plot the standard error for two columns centred at 6 au and 9 au at different times. Note the column near 6 au spans 5.97--6.03 au and  0--$\pi/100$ in the reference state. We observe a fast growth of the column's dispersion.

The expected growth rate of the parametric instability is about $0.32S$, where $S$ is the vertical shear rate \citep{Gammie2000}, provided that $S\ll\Omega$. In the linear regime, from equation~(\ref{eq:vr}), we can estimate
\begin{equation}
\frac{S}{\Omega}=\frac{e_1}{\epsilon}.
\end{equation}
This gives $S/\Omega=1.1$ and 2.1 at 6 au and 9 au, so these approximations are not actually valid. Local simulations with strong horizontal shear flow mimicking nonlinear warped discs find smaller instability growth rates than the estimate of $0.32S$ \citep[see table 2 of][]{Gammie2000}. At 6 au, we measure a growth rate of about 0.2$\Omega$ in Fig.~\ref{fig:std}, which is close to the growth rate (0.25$\Omega$) measured in local simulations with $S=\Omega$ \citep{Gammie2000}. At 9 au, the instability develops only slightly faster than at 6 au.

\begin{figure}
  \centering
  \includegraphics[width=0.9\columnwidth]{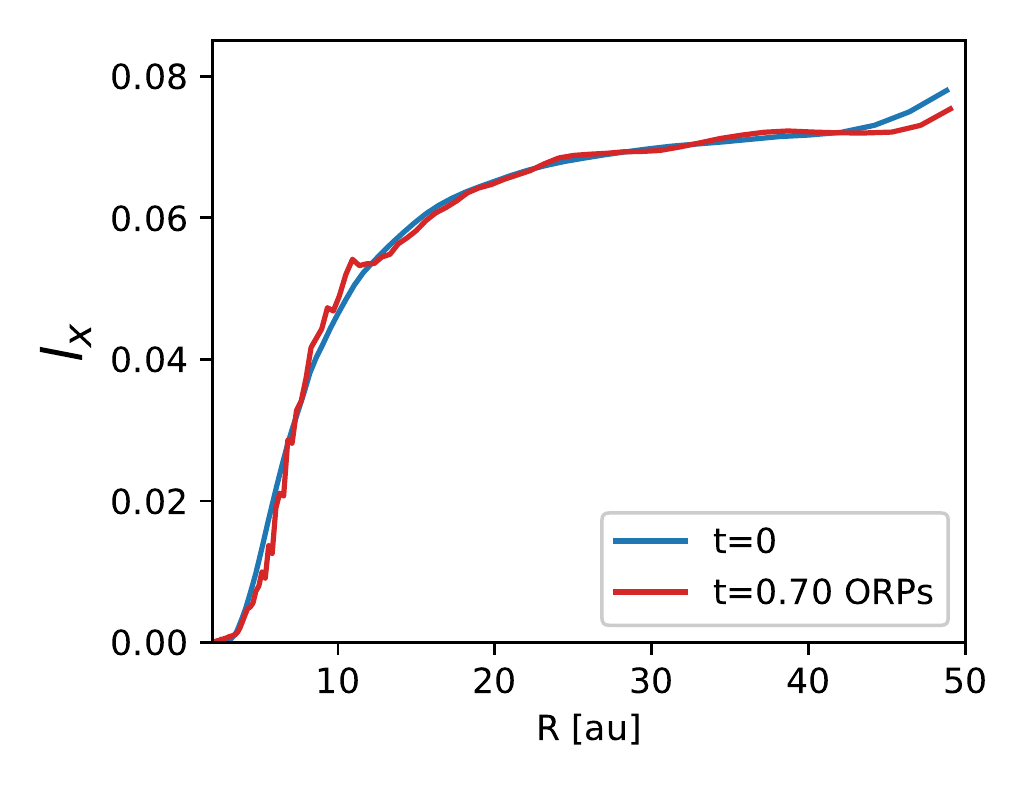}
  \caption{The evolution of $l_x$ in the ultra-high-resolution NS1-MFM-128 simulation. The steady nonlinear warp is altered by the parametric instability.}
 \label{fig:lx128}
\end{figure}

\cite{Deng2021} showed that the parametric instability damps the relative orbital inclination and leads to the decay of free bending waves. However, the warp considered here is maintained by the central quadrupole, as long as it remains misaligned with the outer disc. In Fig.~\ref{fig:lx128}, we check the evolution of the primary tilt vector component, $l_x$. According to Section~\ref{sec:col}, increasing the resolution reduces numerical heating so that the steady warp is expected to be better preserved. At early times, the warp is well preserved in Fig.~\ref{fig:PI}. After the parametric instability grows, the $l_x$ profile is less well preserved than in the low-resolution simulation, NS1-MFM, of Fig.~\ref{fig:lx}. The parametric instability is widespread in low-viscosity warped discs \citep{Ogilvie2013b} but we do not expect it to be decisive in the fast and violent disc breaking process discussed in Section~\ref{sec:break}. The long-term warp profile evolution is yet to be explored.

\begin{figure}
  \centering
  \includegraphics[width=\columnwidth]{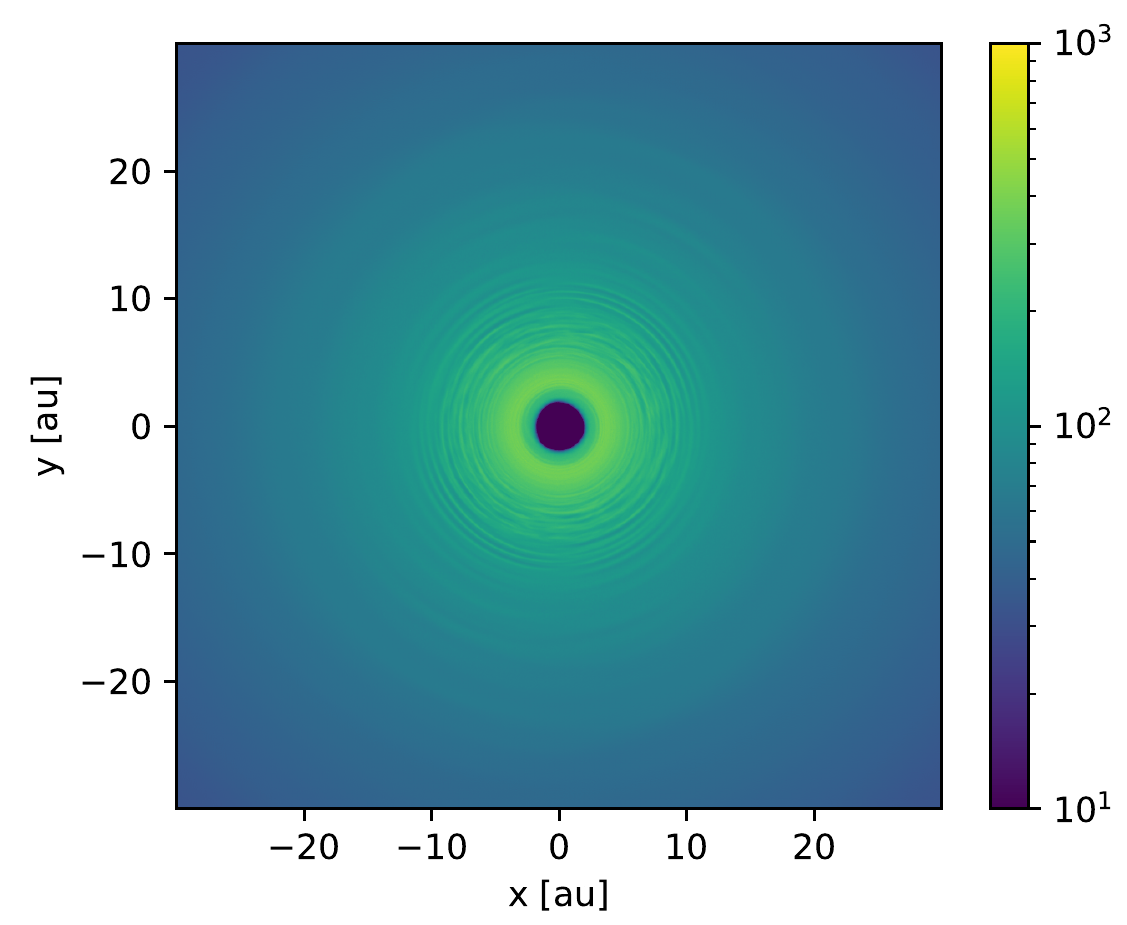}
  \caption{The column density map in cgs units of NS1-MFM-128 at 0.7 ORPs.}
  \label{fig:surf}
\end{figure}

The parametric instability leads to the formation of ring structures, as shown in the column density map in Fig.~\ref{fig:surf}. The rings are more dense, with a characteristic separation of 1 au, in the inner disc while the ring separations appear several times larger in the outer disc. We caution that the simulated time is only about 200 yr in the context of circumstellar discs and is limited by computational resources. However, the rings show no sign of degradation and continue expanding to the outer disc as shown in Fig.~\ref{fig:PI}. In the intriguing observation of HD 139614, multiple rings in the warped disc are revealed by scattered light emission \citep{Muro2020}. In addition, the shadow is too wide to be cast by a broken inner disc \citep[see, e.g.,][]{Facchini2018}. There may be hidden companions in the disc, providing a quadrupolar potential maintaining the warp and rings because free continuous warps tend to decay on a short time-scale \citep{Deng2021}. 

The rings, as local pressure maxima, may trap solids and facilitate dust growth and planet formation. This may have important implications for planet formation around stars with close companions where the secular potential can be approximated as a quadrupole. Disc truncation and wave activity \citep{Artymowicz1994} around a misaligned binary deserve further research.

\section{Conclusions}
\label{sec:con}
We carried out inviscid 3D hydrodynamical simulations of steady warps around a quadrupole with two Lagrangian numerical schemes, MFM and SPH. The steady nonlinear warps are calculated using an affine model \citep{Ogilvie2018}, while the steady solution to the linear bending-wave equations is used as a form of arbitrary initial warp. Our main findings can be summarized as follows:

1. The affine model, in which fluid columns undergo translations and linear transformations, provides a good description of the motion in nonlinear warped discs.

2. MFM follows the fluid column motion well, and thus the warp evolution, when the warp is modest. However, SPH fails to follow the long-term fluid column motion due to excessive numerical dissipation from artificial viscosity. The expansion and compression of the fluid columns are more severe in highly nonlinear warps, providing a challenge for numerical simulations.

3. The affine model predicts a maximum outer tilt of about 0.2 radians in our disc model. Initial warps with smaller outer tilts tend to evolve towards nonlinear steady states, while initial warps with larger outer tilts $>0.2$ break into annuli. In the long term, the annuli reconnect with their neighbours in adiabatic simulations but become isolated in isothermal simulations. SPH produces qualitatively similar results to MFM simulations but with more diffusive substructures. 

4. Parametric instability occurs only in very high-resolution simulations of nonlinear warps, leading to ring structures in the disc. 

\section*{Acknowledgements}

This research was funded by the Isaac Newton Trust (University of Cambridge) through research grant 21.07(d). HD also acknowledges support from the Swiss National Science Foundation via a postdoctoral fellowship, and GIO acknowledges support from STFC through grant ST/T00049X/1. The simulations were performed on the Alps supercomputer of the Swiss National Supercomputing Centre (CSCS). We thank the referee for suggestions that improved the clarity of the manuscript.

\section*{Data Availability}

Data underlying this paper is available from the authors upon reasonable request.



\bibliographystyle{mnras}
\bibliography{references}







\bsp	
\label{lastpage}
\end{document}